 \documentclass[sigconf, screen]{acmart}
\usepackage{float} 

\usepackage{lipsum}  

\usepackage{wrapfig}

\usepackage{booktabs}                  
\usepackage{lipsum}                    
\usepackage{mwe}                       

\usepackage{cellspace}
\usepackage{fontawesome}

\usepackage{enumerate}
\usepackage{enumitem}

\usepackage{makecell} 
\usepackage{booktabs} 

\usepackage{amssymb} 

\usepackage{booktabs} 
\usepackage{tabularx} 
\usepackage[utf8]{inputenc}
\usepackage{geometry}
\usepackage{multicol}

\usepackage{xspace,xpunctuate}

\newcommand{\etal}{{~\textit{et al.}}}

\usepackage{tikz}  
\usetikzlibrary{shadings}
\definecolor{cback}{HTML}{E9ECEF}
\definecolor{cframe}{HTML}{495057}

\newcommand*\circled[1]{\tikz[baseline=(char.base)]{
    \node[shape=rectangle,rounded corners=1.5pt,fill=white,text=black,draw=cframe,inner sep=1pt] (char) {#1};}}

\definecolor{catVisibility}{RGB}{121, 151, 222}     
\definecolor{catSpatial}{RGB}{69, 180, 146}        
\definecolor{catCombination}{RGB}{225, 165, 80}    
\definecolor{catGesture}{RGB}{255, 109, 50}       
\definecolor{catFunction}{RGB}{179, 179, 179}      

\definecolor{catDirect}{RGB}{239, 200, 26}         

\newlength{\boxwidth}
\newcommand{\boxVisibility}{%
  {\setlength{\fboxsep}{1.5pt}%
   \colorbox{catVisibility}{\textsf{\scriptsize\textcolor{white}{\rule{0ex}{0.3ex}\makebox[\boxwidth]{\raisebox{0.15ex}{Ap}}}}}}%
}

\newcommand{\boxMovement}{%
  {\setlength{\fboxsep}{1.5pt}%
   \colorbox{catSpatial}{\textsf{\scriptsize\textcolor{white}{\rule{0ex}{0.3ex}\makebox[\boxwidth]{\raisebox{0.15ex}{M}}}}}}%
}

\newcommand{\boxCombination}{%
  {\setlength{\fboxsep}{1.5pt}%
   \colorbox{catCombination}{\textsf{\scriptsize\textcolor{white}{\rule{0ex}{0.3ex}\makebox[\boxwidth]{\raisebox{0.15ex}{Ar}}}}}}%
}

\newcommand{\boxGesture}{%
  {\setlength{\fboxsep}{1.5pt}%
   \colorbox{catGesture}{\textsf{\scriptsize\textcolor{white}{\rule{0ex}{0.3ex}\makebox[\boxwidth]{\raisebox{0.15ex}{G}}}}}}%
}

\newcommand{\boxFunction}{%
  {\setlength{\fboxsep}{1.5pt}%
   \colorbox{catFunction}{\textsf{\scriptsize\textcolor{white}{\rule{0ex}{0.3ex}\makebox[\boxwidth]{\raisebox{0.15ex}{Af}}}}}}%
}

\newcommand{\boxDirect}{%
  {\setlength{\fboxsep}{1.5pt}%
   \colorbox{catDirect}{\textsf{\scriptsize\textcolor{white}{\rule{0ex}{0.3ex}\makebox[\boxwidth]{\raisebox{0.15ex}{V}}}}}}%
}

\newcommand{\ac}[1]{#1}
\newcommand{\rc}[1]{#1}

\usepackage[nameinlink,capitalise]{cleveref}
\usepackage{array}

\AtBeginDocument{%
  }

\copyrightyear{2025}
\acmYear{2025}
\setcopyright{acmlicensed}\acmConference[UIST '25]{The 38th Annual ACM Symposium on User Interface Software and Technology}{September 28-October 1, 2025}{Busan, Republic of Korea}
\acmBooktitle{The 38th Annual ACM Symposium on User Interface Software and Technology (UIST '25), September 28-October 1, 2025, Busan, Republic of Korea}
\acmDOI{10.1145/3746059.3747678}
\acmISBN{979-8-4007-2037-6/2025/09}

\begin{document}

\title{\textit{InSituTale}: Enhancing Augmented Data Storytelling with Physical Objects}

\author{Kentaro Takahira}
\orcid{0009-0003-5613-610X}
\affiliation{%
  \institution{The Hong Kong University of Science and Technology}
  \city{Hong Kong}
  \country{China}
}
\email{ktakahira@connect.ust.hk}

\author{Yue Yu}
\orcid{0000-0002-9302-0793}
\affiliation{
  \institution{The Hong Kong University of Science and Technology}
  \city{Hong Kong}
  \country{China}
}
\email{yue.yu@connect.ust.hk}

\author{Takanori Fujiwara}
\orcid{0000-0002-6382-2752}
\affiliation{
  \institution{Linköping University}
  \city{Linköping}
  \country{Sweden}
}
\affiliation{
  \institution{University of Arizona}
  \city{Tucson}
  \country{USA}
}

\email{tfujiwara@ucdavis.edu}

\author{Ryo Suzuki}
\orcid{0000-0003-3294-9555}
\affiliation{
  \institution{University of Colorado Boulder}
  \city{Boulder}
  \country{USA}
}
\email{ryo.suzuki@colorado.edu}

\author{Huamin Qu}
\orcid{0000-0002-3344-9694}
\affiliation{%
  \institution{The Hong Kong University of Science and Technology}
  \city{Hong Kong}
  \country{China}
}
\email{huamin@cse.ust.hk}

\renewcommand{\shortauthors}{Takahira et al.}

\crefname{section}{Sec.}{Sec.}


\begin{abstract}
Augmented data storytelling enhances narrative delivery by integrating visualizations with physical environments and presenter actions. Existing systems predominantly rely on body gestures or speech to control visualizations, leaving interactions with physical objects largely underexplored.
We introduce \textit{augmented physical data storytelling}, an approach enabling presenters to manipulate visualizations through physical object interactions.
To inform this approach, we first conducted a survey of data-driven presentations to identify common visualization commands. 
We then conducted workshops with nine HCI/VIS researchers to collect mappings between physical manipulations and these commands.
Guided by these insights, we developed \textit{InSituTale}, a prototype 
that combines object tracking via a depth camera with Vision-LLM for detecting real-world events.
Through physical manipulations, presenters can dynamically execute various visualization commands, delivering cohesive data storytelling experiences that blend physical and digital elements.
A user study with 12 participants demonstrated that \textit{InSituTale} enables intuitive interactions, offers high utility, and facilitates an engaging presentation experience.
\end{abstract}

\ccsdesc[500]{Human-centered computing~Visualization design and evaluation methods}

\keywords{Visualization; Data-Driven Storytelling; Tangible Interaction; Augmented Reality; Augmented Presentation; Video}

\begin{teaserfigure}
   \includegraphics[width=\linewidth]{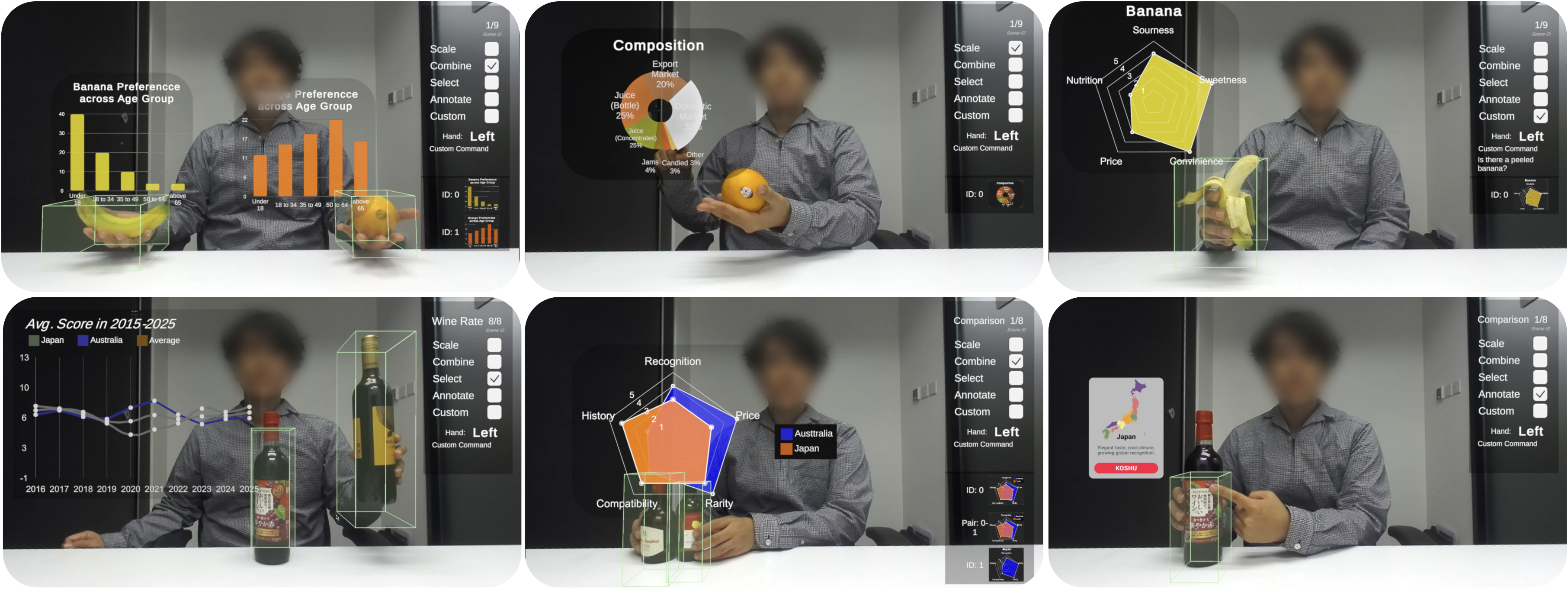}
  \caption{\textit{InSituTale}: We developed \textit{InSituTale}, an augmented physical data storytelling prototype to enable presenters to control visualizations through physical object manipulations, achieving coordination of physical and digital elements.}
  \label{fig:teaser}
  \Description{Figure 1.1
A composite image showing six examples of using the InSituTale system for augmented physical data storytelling. (1) A fruit is detected using an object detection algorithm, triggering a chart to appear. (2) A physical object is moved closer to another chart, causing the charts to combine. (3) The object is lifted to select a line in a line chart. (4) The object points at a chart to highlight a specific data point. (5) A pointing gesture reveals an annotation tied to the object. (6) A banana is peeled, causing a chart to change.}
\end{teaserfigure}

\maketitle

\section{Introduction}
Augmented data storytelling enriches data-driven narratives by integrating visualizations with physical environments and presenter actions.
Recent systems have demonstrated the potential of controlling visualizations through body gestures~\cite{chironomia, visConductor, BodyDrienGraphics, hanstreamer} and speech~\cite{srinivasancombining, realityTalk}, enhancing performative aspects and enabling improvisational storytelling experiences that respond dynamically to audience engagement.
However, these systems predominantly focus on gesture- and speech-driven interactions, overlooking the role of physical objects as integral storytelling components or interaction mediums.

This limited focus represents a notable gap. In real-world data storytelling, such as product demonstrations or educational contexts, physical objects often serve as referents for visualizations, playing a central role in conveying data narratives~\cite{ChemiAR, DataStoryHybrid, ProductDemoAR}.
The coordinated manipulation of physical objects and associated visualizations can enhance audience understanding and engagement, as illustrated in professionally edited augmented videos~\cite{adidas, rosling2016WhatIf}.
Despite the rich affordances of physical objects, including grasping, rotating, combining, and transforming, existing augmented data storytelling systems offer limited support for these interactions~\cite{chironomia, hanstreamer, realityTalk, realityEffect, realityTalk}. Consequently, current solutions often fail to deliver cohesive storytelling experiences that effectively blend physical and digital elements.

In this paper, we introduce \textit{augmented physical data storytelling}, a novel approach enabling presenters to intuitively control visualizations through direct manipulations of physical objects. 
We expect that coupling physical object interactions meaningfully with visualization responses can offer presenters intuitive control and expressive storytelling capabilities, seamlessly blending physical and digital elements.
Additionally, tangible interactions leveraging rich physical affordances are expected to reduce presenters' cognitive load and offer more expressive interactions than purely gesture-based methods~\cite{TangibleNet}.

To inform our approach, we first conducted a survey of data-driven presentations to identify a relevant set of visualization commands (e.g., showing charts, selecting data points, and changing chart types). Subsequently, we held workshops with nine HCI/VIS researchers to identify intuitive mappings between physical manipulations and these visualization commands. 
The workshops produced diverse user-generated mappings categorized into six groups: 1) appearance-based, 2) movement-based, 3) arrangement-based, 4) gesture-based, 5) affordance-based, and 6) visualization-based interactions. 
We analyzed these ideas and identified common mappings for each command.
Drawing upon these insights, we iteratively developed \textit{InSituTale}, a prototype supporting augmented physical data storytelling. This process also yielded five critical design considerations: 1) physical space interaction detection, 2) dynamic visualization placement, 3) minimization of interaction ambiguity, 4) smooth storytelling flow, and 5) context-aware presenter guidance.

\textit{InSituTale} consists of two primary modes: presentation and authoring. In the presentation mode, presenters control visualizations in real time using physical object manipulations. A depth camera captures physical manipulations, including pointing gestures, lifting objects, adjusting distances between objects, and moving objects closer to or farther from the camera. 
Additionally, a vision-language model supports real-time detection of custom-defined changes in object states (e.g., \textit{``Is the banana peeled?''}), which can trigger corresponding visualization updates. These capabilities enable presenters to execute a wide range of visualization commands, including showing/hiding charts, scaling, composing or decomposing charts, selecting individual data points or series, switching chart types, and transitioning between overview and detailed views. In the authoring mode, presenters configure scenes by assigning visualizations to physical objects, adding annotations, and defining customizable commands. 

We evaluated \textit{InSituTale} with 12 participants, assessing its usability, utility, and learnability. Quantitative and qualitative results confirmed that \textit{InSituTale} supports intuitive interactions, provides high utility, and facilitates engaging presentation experiences.
This paper makes the following main contributions:
\begin{itemize}[noitemsep,topsep=0pt,label=$\diamond$, leftmargin=*]
\item  \textbf{Novel Concept and Interaction Design}: 
We introduce \textit{augmented physical data storytelling} and present insights into intuitive mappings between physical object manipulations and visualization commands through design workshops. 
\item  \textbf{System Design and Implementation}: We developed \textit{InSituTale}, a prototype that integrates object tracking and a vision-language model to support real-time, physically driven interactions with visualizations.
\item \textbf{User Study}: 
We report on user studies highlighting system strengths, limitations, and implications for designing future augmented data storytelling systems.
\end{itemize}
\section{Related work}
Our research examines the intersection of data-driven storytelling, augmented presentation, and physical object-based interactions for data visualization. We review these domains and illustrate how our work complements prior studies.

\label{sec:related_work}
\subsection{Data-Driven Storytelling}

Data-driven storytelling~\cite{LeeSketchStory, FromJamtoRecital}, where presenters deliver real-time, data-driven narratives, is a valuable practice in various contexts, such as organizational decision-making and public communication~\cite{FromJamtoRecital}. In particular, improvisational storytelling, characterized by small audiences and spontaneous data communication, known as \textit{jam-session-style}~\cite{FromJamtoRecital}, requires presenters to adapt their narratives in response to audience interactions, creating personalized and engaging presentation experiences by interacting with visualizations. 
Systems designed to support interaction in data-driven storytelling have been proposed, from AR-based~\cite{chironomia, visConductor, TangibleNet} to screen-based presentations~\cite{LeeSketchStory, FromJamtoRecital, srinivasancombining}. These systems focus on facilitating easy and direct manipulation of visualizations rather than at the slide level~\cite{FromJamtoRecital, LeeSketchStory}. 
Also, they often utilize gestures or speech to enhance the performative quality of the presentations~\cite{chironomia, srinivasancombining, TangibleNet}.

In jam-session-style scenarios, physical objects can play a pivotal role by enriching the storytelling experience~\cite{Tangible3DTableTop, riche2018data}. 
As props, these objects serve as the central focus of the presentation and tangible anchors for abstract data, helping bridge the gap between complex information and audience comprehension.
For instance, product demonstrations often pair real physical products with visualizations displayed on separate screens or papers~\cite{ProductDemoAR}. 
Similarly, Hans Rosling's iconic presentations showcased the powerful use of physical objects like boxes and stones, which he manipulated by stacking, hiding, and relocating to illustrate demographic dynamics~\cite{rosling2016global,rosling2007best, rosling2016world, rosling2014global}. These techniques contextualize abstract data with physical objects, making the information more tangible and understandable while also boosting audience engagement. 
The theatrical theory further underscores the performative quality of props, highlighting their ability to captivate audiences and strengthen communication~\cite{crowther2019theory}. Despite these compelling examples, the use of physical objects in data storytelling remains largely overlooked~\cite{riche2018data, TangibleNet}.
In this paper, we fill this gap by introducing augmented physical data storytelling.
\subsection{Augmented Presentation}
Augmented presentations~\cite{realityTalk, BodyDrienGraphics}, which overlay digital content onto presenters and their physical environment, are gaining popularity across domains such as education~\cite{realityTalk, surpriseMachine}, advertising~\cite{ARProductPresentation}, and business presentation~\cite{chironomia}. 
Existing research has explored various approaches to augmenting presentations~\cite{BodyDrienGraphics, hanstreamer, PoseTween, chironomia, realityTalk, Elastica, realityEffect, surpriseMachine, MultimodalDirectManipulation, Chalktalk}. Saquib\etal~\cite{BodyDrienGraphics} introduced body-driven graphics that map visualizations to specific body parts, dynamically adapting to presenters’ movements. Hall\etal~\cite{chironomia} developed Augmented Chironomia, a system that enables gesture-based visualization control for remote presentations. RealityTalk~\cite{realityTalk} employs a keyword-matching system to recognize spoken words and generate corresponding graphic elements that presenters can manipulate through gestures. Liu\etal~\cite{PoseTween} proposed PoseTween, an authoring tool that animates objects based on human poses, ensuring natural coordination between human actions and object animations. Elastica~\cite{Elastica} addresses challenges in recognition errors and presenter mistakes by allowing dynamic adjustments to predefined graphic animations through speech and gestures. RealityEffects~\cite{realityEffect} augments volumetric 3D scenes, enabling users to bind captured physical objects with annotated visual effects that dynamically respond to physical motion.

While these approaches offer various modalities in different ways, they offer limited coordination between physical objects and visualizations. Most existing systems either overlook the role of physical objects~\cite{Elastica, chironomia, BodyDrienGraphics, PoseTween} or focus on basic interactions, such as tracking objects and having visuals follow their movements~\cite{realityTalk, realityEffect}. 
Consequently, they fail to fully utilize the diverse manipulations that physical objects afford, such as stacking, rotating, colliding, and bringing multiple objects closer together. 
\ac{By contrast, traditional video-editing tools, while capable of incorporating physical props, lack the real-time flexibility required for improvisational storytelling, particularly in mid- to small-scale settings~\cite{FromJamtoRecital}.}
Our work addresses this gap by designing augmented presentations that facilitate effective coordination between physical objects and visualizations, with a particular focus on data storytelling. 

\subsection{Interacting with Visualization Using Physical Objects}
Interacting with visualizations through physical objects offers substantial benefits, including reduced cognitive load and learning cost~\cite{TangibleBit, TUItextbook}. Physical objects provide an intuitive and direct way to engage with data, leveraging users’ natural spatial reasoning and motor skills to minimize the cognitive effort required for interaction~\cite{IshiiTUI}. 
To further bridge the gap between the virtual and physical realms, embedded data representations integrate visualizations with the physical objects or spaces to which it refers, known as referents~\cite{SituatedAnalyticsTextbook,embeddedDataRep, SituatedAnalytics}. 
Techniques for integrating visualizations with physical referents include adding labels\cite{ARLabelZhutian, LabellingOutOfView}, overlaying information~\cite{motionVis, Pearl}, nearby placement~\cite{EffectsOfViewLayout, AugmentedObjectIntelligence}, and spatial projections~\cite{VisTorch}. 
These methods create a seamless connection between data and its physical referents.

When physical objects are used as referents, their properties---such as being pickable, stackable, or combinable---together with their semantic relationships to visualizations, open up diverse and meaningful opportunities for interacting with visualizations~\cite{PaperVis, activeProxyDash, urp}. 
For instance, Uplift~\cite{uplift} allows users to display building information by picking up the corresponding scale models from a table. 
Satriadi\etal~\cite{tangibleGlobe} explored tangible globes for geospatial visualizations, enabling physical manipulation of embedded maps and data. 
Additionally, Active Proxy Dashboard~\cite{activeProxyDash} uses scale models to interact with dashboards, supporting operations like filtering data by picking up specific models or authoring composite visualizations by bringing multiple models together. 
Herman\etal~\cite{MultiTouchDataPhy} introduced landscape models that integrate geospatial data and simulations, allowing users to adjust cutting planes and simulations precisely within the same spatial coordinate system as the physical model.

While these studies provide valuable insights into interaction design with physical objects in augmented environments, most focus on analytics purposes, where the primary goal is to support data exploration~\cite{SituatedAnalytics}.
In contrast, data storytelling introduces unique requirements, such as the performative aspects of interaction~\cite{chironomia}, real-time responsiveness to audience discussions~\cite{FromJamtoRecital}, and intuitive, low-cost interactions that allow presenters to focus on delivering their narrative without errors~\cite{LeeSketchStory, Elastica}.  
Studies on data physicalization further highlight the communicative potential of tangible artifacts to make abstract data more tangible and engaging~\cite{NetworkPhyAuthor, DataPhyForTechComm, InterPersonalDataNarratives}. However, many of these approaches remain static, constrained by technical limitations that prevent dynamic visualization changes or real-time interactions---features critical for data storytelling. 
Our work contributes to this field by designing an interaction framework for real-time data storytelling, leveraging physical objects as referents. 
\section{Augmented Physical Data Storytelling}
This section introduces the concept of \textit{augmented physical data storytelling} along with its key features.

\subsection{Concept}
We propose \textit{augmented physical data storytelling},
An approach that integrates physical object manipulations with data visualizations to deliver engaging, real-time narratives.
This approach introduces an interface that synchronizes the movement and transformation of physical objects with corresponding changes in visualizations, blending physical artifacts, digital visuals, and presenter gestures into a cohesive mixed-reality experience.
Unlike prior methods that rely on extensive post-production, our approach enables seamless, live presentation without the need for video editing.
\ac{Furthermore, unlike previous studies that use physical objects for visualization interactions, our approach is designed to support novel interaction techniques tailored to real-time data storytelling contexts.}

Our approach draws inspiration from the presentation style of Hans Rosling, who captivated audiences by integrating multiple modalities---gestures, speech, physical objects, and visuals---to create an engaging and cohesive storytelling experience. 
Rosling's presentations often involved post-edited animations on visualizations and synchronized them with his gestures to simulate direct control over the visuals~\cite{rosling2013Myth, rosling2016WhatIf, rosling2013DontPanic}. He also used tangible props like boxes, paper rolls, and stones to make abstract demographic data more relatable~\cite{rosling2014global,rosling2016global,rosling2016world}.

Inspired by this interplay between the physical and the visual, our goal is to realize a real-time, end-to-end system that preserves this expressiveness while enabling real-time storytelling. We envision applications across domains such as education, product demonstrations, and public presentations, where physical objects serve as essential narrative props, and their coordination with visualizations enhances the clarity and communicative power of data-driven narratives.

\subsection{Key Features}
To realize the concept of \textit{augmented physical data storytelling}, the system incorporates the following core features:

\subsubsection*{\textbf{Coordination of Physical Objects and Visualizations}}
The system enables real-time synchronization between physical object manipulations and corresponding changes in visualizations. Movements such as repositioning, arranging, or transforming objects dynamically trigger visual updates, fostering a seamless and intuitive link between the physical and digital elements. 
\ac{Our approach supports a range of semantic couplings, from loosely performative to tightly meaningful. For example, lifting a physical object may simply serve to reveal a related visualization in a performative manner. In contrast, changing the state of a physical object, such as peeling a banana, can trigger visual updates that reflect semantic meaning, such as nutritional information tied to the object's edible state. }

\subsubsection*{\textbf{Support for Diverse Interactions}}
To meet the expressive needs of data storytelling, the system supports a wide range of interaction types with visualizations~\cite{chironomia, TangibleNet, LeeSketchStory}. This flexibility enables presenters to construct rich narratives that adapt to varying data complexities and audience engagement.

\subsubsection*{\textbf{Improvisational Presentation}}
The system is designed to support non-linear, adaptive storytelling, as emphasized in prior work~\cite{FromJamtoRecital}. Presenters can adjust the narrative flow in real time based on audience reactions or spontaneous insights, enabling more engaging and personalized presentations.
\section{Design Workshops to Solicit Interactions}
\label{workshop}
Through a formative workshop, we explored how physical object manipulations can naturally coordinate with visualization commands used in data storytelling. 
For instance, when introducing a new product, a presenter may pick it up, point to it, move it closer to the audience, or adjust its shape to emphasize particular features. 
Suppose such physical manipulations can be meaningfully mapped to visualization commands through well-designed interaction mechanisms.
In that case, it raises the question of which mappings between physical object manipulations and visualization commands are intuitive and effective for storytelling. To answer this question, we conducted design workshops to identify effective mappings that could inform the design of augmented physical data storytelling systems.

\subsection{Visualization Commands for Storytelling}
Our first goal was to identify a relevant set of visualization commands for data storytelling, such as showing, hiding, and scaling visualizations. Given the wide range of possible commands, we focused on those most relevant to real-time presentation settings. 
To ensure that the selected commands reflect authentic storytelling practices, we conducted an exploratory survey of publicly available videos showcasing augmented or interactive presentations with data visualizations. 
We performed keyword-based searches on platforms such as YouTube and Vimeo using terms including ``data-driven storytelling,'' ``AR presentation,'' ``virtual presentation,'' and ``data presentation.''
Due to the absence of standardized repositories or consistent keywords, we adopted a manual, iterative collection strategy. 
\ac{Rather than starting from a broader pool, we directly collected matching examples based on their relevance to our goals.}
We selected videos that featured a presenter delivering data-driven content to an audience,  using visualizations in conjunction with physical or gestural interactions. 
This process resulted in a curated dataset of 31 videos (see the supplemental materials).
While not exhaustive, this dataset served as a resource for identifying common visualization commands used in practice.
Three researchers independently analyzed this dataset and then collaboratively discussed their observations to reach a consensus. Through this process, we distilled a set of key visualization commands, including: 


\begin{wrapfigure}[2]{l}{0.001\textwidth}
  \begin{center}
    \vspace{-15pt}
    \includegraphics[width=0.035\textwidth]{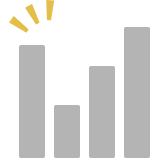}
  \end{center}
\end{wrapfigure}
\noindent\textbf{Show/Hide}: Toggling the visibility of a visualization during the presentation. This command appeared in nearly all the videos, as it serves as a fundamental mechanism for controlling the narrative flow.

\begin{wrapfigure}[2]{l}{0.001\textwidth}
  \begin{center}
    \vspace{-15pt}
    \includegraphics[width=0.035\textwidth]{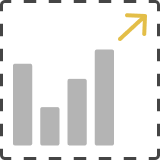}
  \end{center}
\end{wrapfigure}
\noindent\textbf{Scale:} Adjusting the size of a visualization to control focus and legibility. This command was commonly used to draw the audience's attention to a specific chart or element, especially when emphasizing the details (\cref{fig:DataVideoExample}-A).

\begin{wrapfigure}[2]{l}{0.001\textwidth}
  \begin{center}
    \vspace{-15pt}
    \includegraphics[width=0.035\textwidth]{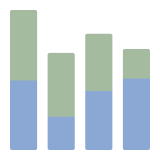}
  \end{center}
\end{wrapfigure}
\noindent\textbf{Compose/Decompose:} Merging multiple visualizations into a unified view or separating them into distinct parts.
For instance, overlaying two bar charts can help compare two products.
This technique was often used to emphasize part-whole relationships or contrast different data sources (\cref{fig:DataVideoExample}-B).

\begin{wrapfigure}[2]{l}{0.001\textwidth}
  \begin{center}
    \vspace{-15pt}
    \includegraphics[width=0.035\textwidth]{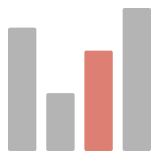}
  \end{center}
\end{wrapfigure}
\noindent\textbf{Select/Deselect Data Points:}
Highlighting an individual data point. This command was often used to draw attention to a specific value during narration (\cref{fig:DataVideoExample}-C).

\begin{wrapfigure}[2]{l}{0.001\textwidth}
  \begin{center}
    \vspace{-15pt}
    \includegraphics[width=0.035\textwidth]{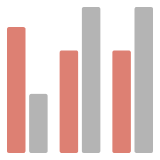}
  \end{center}
\end{wrapfigure}
\noindent\textbf{Select/Deselect Data Series:} Selecting a group of data points that share the same category (e.g., all countries in Asia). This was commonly used to highlight patterns or differences across categories (\cref{fig:DataVideoExample}-D).

\begin{wrapfigure}[2]{l}{0.001\textwidth}
  \begin{center}
    \vspace{-15pt}
    \includegraphics[width=0.035\textwidth]{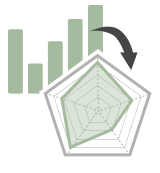}
  \end{center}
\end{wrapfigure}
\noindent\textbf{Change Chart Types:}
Switching between chart types (e.g., from a bar chart to a line chart) to convey different aspects of the same data source. This command was used to show multiple perspectives on the data (\cref{fig:DataVideoExample}-E).

\begin{wrapfigure}[2]{l}{0.001\textwidth}
  \begin{center}
    \vspace{-15pt}
    \includegraphics[width=0.035\textwidth]{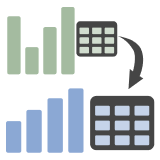}
  \end{center}
\end{wrapfigure}
\noindent\textbf{Change Data Sources:} Switching data sources being visualized to align with the narrative context or the state of the physical object. This can be done alongside a change in the visualization type.

\begin{wrapfigure}[2]{l}{0.001\textwidth}
  \begin{center}
    \vspace{-15pt}
    \includegraphics[width=0.035\textwidth]{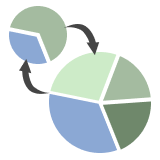}
  \end{center}
\end{wrapfigure}
\noindent\textbf{Hierarchical Navigation:}
Navigating between overview and detail views within a visualization. This commonly followed a ``zoom-in'' narrative structure---for example, presenting a broad category using a pie chart and then revealing its breakdown into subcategories (\cref{fig:DataVideoExample}-F).

\begin{figure}[t]
    \centering
    \includegraphics[width=1\linewidth]{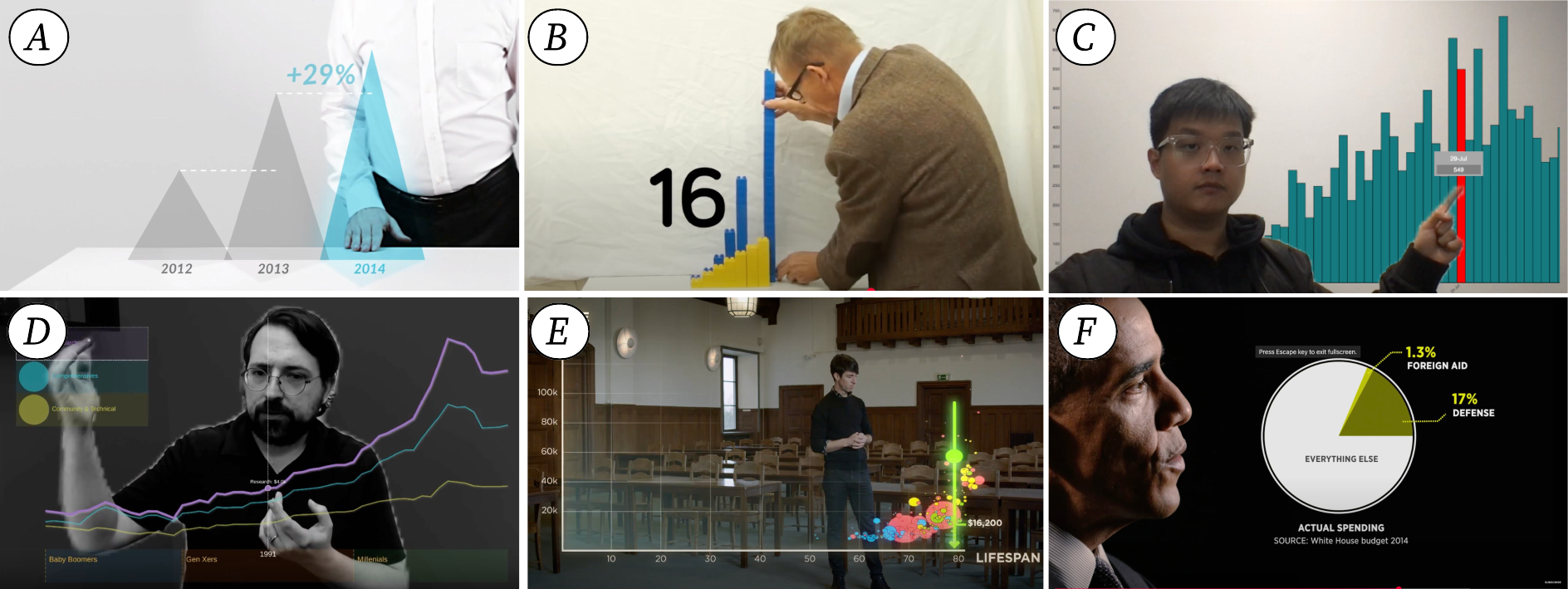}
     \caption{Interactions in Data-Driven Presentations: (A) scale up~\cite{VisualAgencyVideo}, (B) overlay bar charts~\cite{EbolaStoppedNoe}, (C) select a bar~\cite{hanstreamer} (D) select data series~\cite{FlowImmersive}, (E) change a bubble chart to a bar chart~\cite{betterAndBetter}, (F) show a detailed view~\cite{Obama}.}
    \label{fig:DataVideoExample}
    \Description{Figure 2.1
Six examples of interactions in data-driven presentations taken from online videos. (A) A bar chart is scaled up for emphasis. (B) Bar charts are overlaid on a background image. (C) A user selects a specific bar from a chart. (D) A data series is selected interactively. (E) A bubble chart transitions into a bar chart. (F) A detailed view is displayed on demand.}
\end{figure}

\subsection{Selection of Physical Objects}
To ground discussions during the workshop, we prepared a set of physical objects to explore interaction mappings. 
Guided by prior work on object affordances~\cite{AffordanceBasedTangible}, we focused on three key dimensions that likely affect users' physical object manipulations: size, grabbability, and edge characteristics. 
We selected tabletop-sized objects to reflect typical use contexts---such as product demonstrations---and to support diverse hand-scale interactions.
Based on these considerations, we selected a set of objects spanning a range of dimensions, including a cup, wine bottle, banana, toy car, backpack, and laptop (\cref{tab:workshop}). 
 
\vspace{0pt} 
\begin{table}[h]
    \caption{Selected Physical Objects: We provided the six objects to cover diverse ranges of physical properties.}
    \label{tab:workshop}
     \Description{
Table 1.1 : A table summarizing six selected physical objects used in the study—Cup, Bottle, Banana, Toy Car, Backpack, and Laptop—each shown with an image. The table includes their size (S, M, L), whether they are easily grabbable (check or cross), and whether they have curved edges (check or cross). For example, the cup is small, grabbable, and has curved edges, while the laptop is medium-sized, not grabbable, and has no curved edges.
     }
    \centering
    \small
    \setlength{\tabcolsep}{3pt}
    \begin{tabular}{ccccccc}
     & Cup & Bottle & Banana & Toy Car & Backpack & Laptop \\
     & 
    \multicolumn{1}{c}{\includegraphics[width=0.04\textwidth]{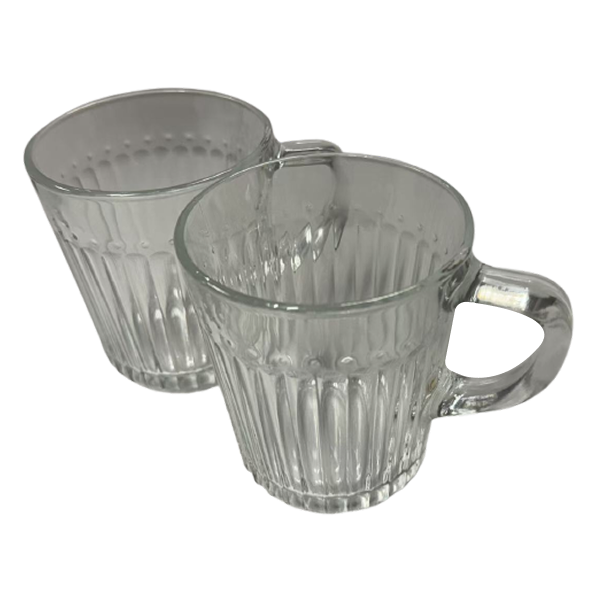}} & 
    \multicolumn{1}{c}{\includegraphics[width=0.04\textwidth]{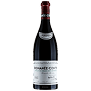}} & 
    \multicolumn{1}{c}{\includegraphics[width=0.04\textwidth]{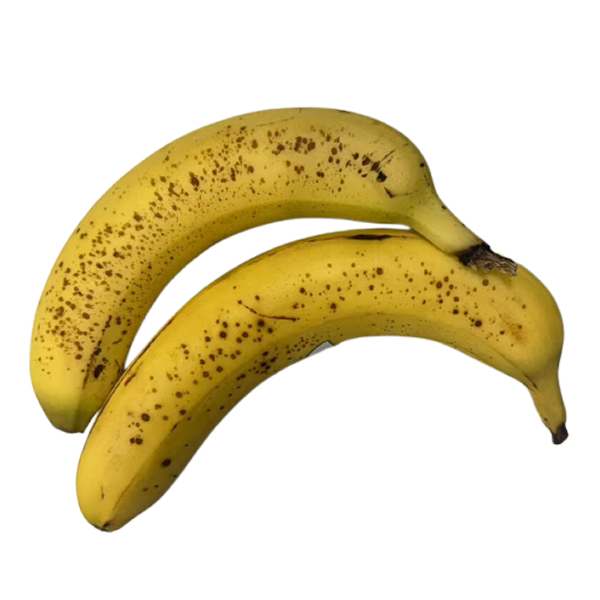}} & 
    \multicolumn{1}{c}{\includegraphics[width=0.04\textwidth]{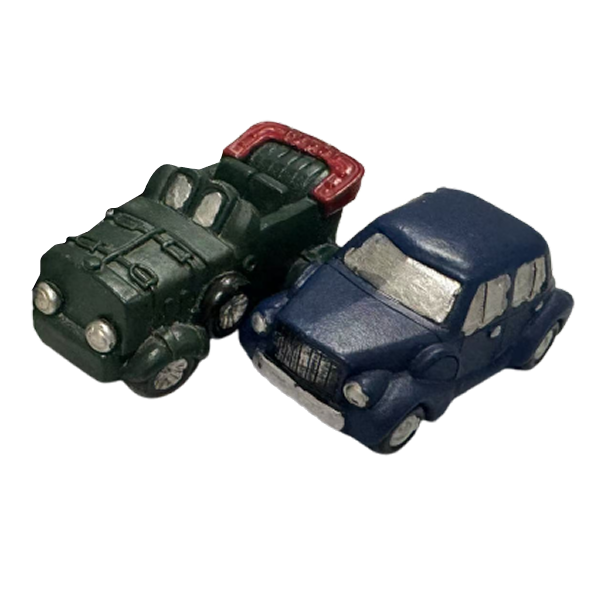}} & 
    \multicolumn{1}{c}{\includegraphics[width=0.04\textwidth]{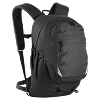}} & 
    \multicolumn{1}{c}{\includegraphics[width=0.04\textwidth]{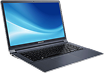}} \\
    \hline
    \textbf{Size} & \textbf{S} & \textbf{M} & \textbf{M} & \textbf{S} & \textbf{L} & \textbf{M} \\
    \hline
    \textbf{Grabbable} & \faCheck & \faCheck & \faCheck & \faCheck & \faClose & \faClose \\
    \hline
    \makecell{\textbf{Curved} \\ \textbf{Edge}} & \faCheck & \faCheck & \faCheck & \faClose & \faCheck & \faClose \\
    \hline
    \end{tabular}
\end{table}
\vspace{-5pt} 

\subsection{Ideation Workshop}
The ideation workshop involved researchers in structured activities, where they engaged in brainstorming and group discussions to develop a variety of interaction ideas. 
Three sessions were held, each with three participants. Sessions continued until the proposed mappings reached saturation\ac{; while saturation wasn’t formally measured, 90\% of the final participant’s mappings in the last workshop overlapped with earlier ones.}

\begin{figure*}[ht]
    \centering
    \includegraphics[width=1\linewidth]{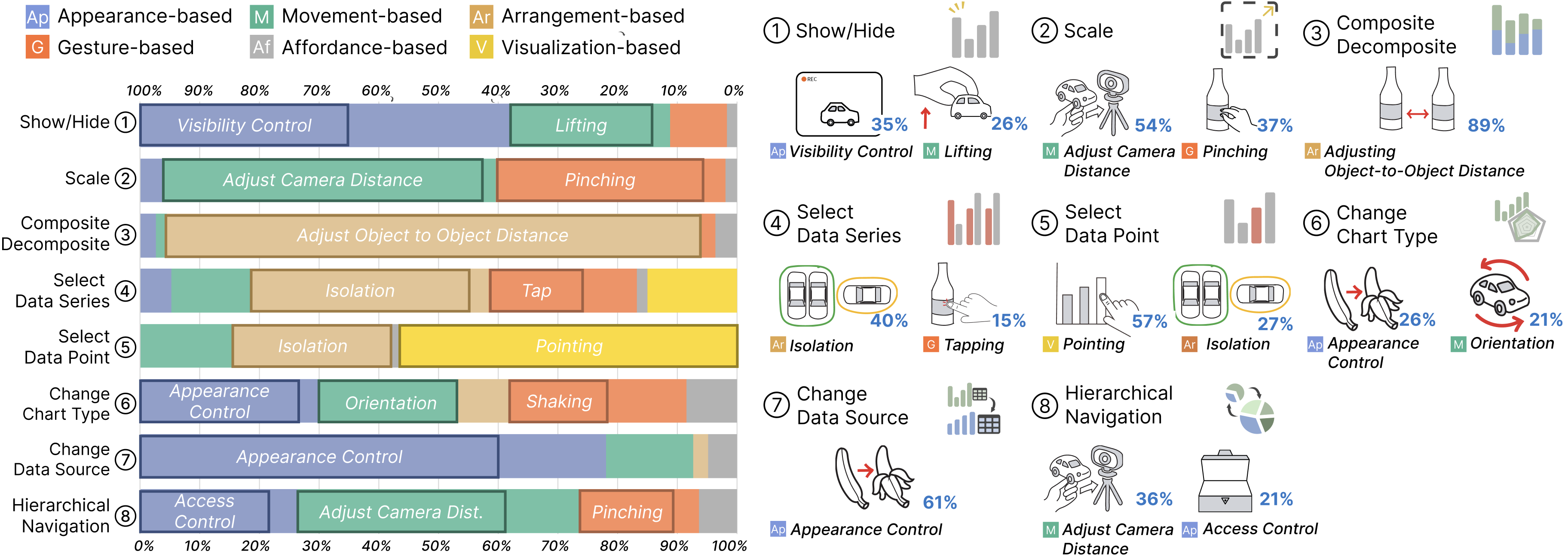}
    \caption{Workshop Results: This illustrates mappings between physical manipulations and visualization commands across various physical objects. 15 types of manipulations are grouped into six categories. For each visualization command, we present the cumulative percentage of manipulations within each category, along with the top two most frequently proposed manipulations across categories. A detailed breakdown of individual objects is provided in the appendix.}
    \label{fig:workshopresult}
     \Description{Figure 3.1: This figure presents a bar chart summarizing the results of a workshop that investigated how users mapped physical actions to interaction commands using various physical objects. Fifteen unique user actions are grouped into six higher-level categories. For each interaction command, the chart displays the cumulative percentage of proposed user actions falling into each category, using stacked bars to indicate the distribution. The figure also highlights, for each command, the top two most frequently mentioned specific user actions, which are labeled across the chart. These mappings reflect common physical gestures or manipulations—such as tapping, lifting, or rotating objects—that participants associated with particular interactive behaviors. A more detailed object-level breakdown is available in the appendix.}
\end{figure*}

\subsubsection{Participants}
Our workshop involved nine participants (W1--W9), including researchers specializing in Visualization (VIS) and Human-Computer Interaction (HCI). \ac{Participants (four HCI, five VIS) had 4–10 years of experience and} selected for their expertise in interaction design, following methodologies in similar studies~\cite{DataCube, PaperVis, TangibleNet}. The participants ranged in age from 23 to 32 and included seven males and two females. 


\subsubsection{Materials}
Participants received a set of visualization commands with accompanying illustrations to support comprehension (see supplemental materials). These materials remained accessible throughout the workshop. Participants were also provided with the six selected physical objects and encouraged to experiment with potential user manipulations for the given commands. To facilitate idea generation, we introduced several example mappings, which served as a priming technique~\cite{primingTech}, commonly used in previous studies~\cite{DataCube, PaperVis, TangibleNet}, to stimulate creative thinking and clarify possible interaction mappings.

\subsubsection{Procedure}
Each 70-minute session comprised three phases: briefing, individual brainstorming, and group discussion. In the briefing, participants completed a consent form, followed by a 10-minute introduction covering the research background and workshop objectives. We explained the visualization commands and demonstrated the example mappings while emphasizing that these were merely examples and encouraging creative thinking beyond them. During individual brainstorming (25--30 minutes), participants explored ways to perform each visualization command by manipulating the provided physical objects.
The participants were asked to record their mapping ideas on worksheets. They were also encouraged to think freely without concern for technical limitations. In the group discussion (20--30 minutes), participants demonstrated their ideas using the physical objects, discussed possible extensions, and explained the reasoning behind each mapping. The entire discussion was recorded and documented in the instructor's notes for later analysis.

\subsubsection{Data Analysis}
Two authors organized the workshop outcomes, identifying a total of 143 unique manipulations across the six physical objects. 
Proposed manipulations were initially coded using a reference set derived from prior studies~\cite{ TangibleTouch, AffordanceBasedTangible, Tangible3DTableTop}, which included common manipulations such as tapping, colliding, and stacking. 
An open-coding approach was applied to manipulations beyond this initial set. Coding was refined iteratively through discussions among two authors until consensus was achieved. 

\subsection{Results}
We summarized the results in terms of prominent physical manipulation categories and frequently proposed mappings.

\subsubsection{Manipulation Categories}
We synthesized the diverse manipulation ideas from participants into 15 distinct types, organized into six overarching categories based on the nature of the manipulations and the intentions.

\noindent\boxVisibility\hspace{0.2em}\textbf{Appearance-based Interactions} control what is exposed to the system or the audience. This includes \textit{visibility control} (e.g., showing or hiding a physical object from the camera's view), \textit{appearance control} (e.g., transforming the shape of a physical object), and \textit{access control} (e.g., revealing or concealing the contents when opening a bag or removing a bottle cap).

\noindent\boxMovement\hspace{0.2em}\textbf{Movement-based Interactions} involve changing a physical object’s position or orientation. The proposed manipulations include \textit{lifting}, \textit{relocating}, \textit{changing orientation}, or \textit{changing the distance from the camera}.

\noindent\boxCombination\hspace{0.2em}\textbf{Arrangement-based Interactions} relate to the spatial arrangement of multiple physical objects. This category includes \textit{adjusting distances among physical objects} and \textit{isolating a specific object} visually or spatially from others (e.g., lifting one while leaving others in place).

\noindent\boxGesture\hspace{0.2em}\textbf{Gesture-based Interactions} are hand gestures performed on or near an object, including \textit{tapping}, \textit{shaking}, \textit{pinching}, and \textit{drawing} on surfaces with a finger.

\noindent\boxFunction\hspace{0.2em}\textbf{Affordance-based Interactions} utilize the inherent affordances of a physical object to trigger commands. Examples include moving a toy car door, tightening a bag strap, drinking a bottle of wine, or pressing a laptop's spacebar.

\noindent\boxDirect\hspace{0.2em}\textbf{Visualization-based Interactions} directly control visualizations without involving physical objects. Participants typically proposed these manipulations when perceived to be more immediate or intuitive than controlling visualizations through physical objects. 
Common examples include \textit{pointing at visuals} and \textit{enclosing them with both hands}.

\subsubsection{Dominant Mappings}
Based on the categorized types of manipulations, we analyzed frequently proposed mappings between physical manipulations and visualization commands (\cref{fig:workshopresult}). 
Refer to our supplementary materials for object-specific analyses.

\begin{wrapfigure}[2]{l}{0.001\textwidth}
  \begin{center}
    \vspace{-15pt}
    \includegraphics[width=0.035\textwidth]{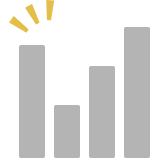}
  \end{center}
\end{wrapfigure}
\noindent\textbf{Show/Hide:} 
Participants frequently mapped this command to \textbf{Appearance-based} and \textbf{Movement-based Interactions}, such as \boxVisibility\hspace{0.1em}\textit{controlling visibility from the camera} (35\%), \boxMovement\hspace{0.1em}\textit{lifting the object} (27\%), and \boxVisibility\hspace{0.1em}\textit{controlling access} (26\%). Controlling visibility or lifting was commonly suggested for small, easily graspable objects, while access controls were favored for heavier or less grabbable items (e.g., opening a backpack).

\begin{wrapfigure}[2]{l}{0.001\textwidth}
  \begin{center}
    \vspace{-15pt}
    \includegraphics[width=0.035\textwidth]{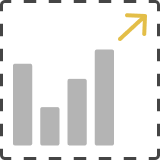}
  \end{center}
\end{wrapfigure}
\noindent\textbf{Scale:}
This command was primarily associated with \textbf{Movement-based} and \textbf{Gesture-based Interactions}, particularly \boxMovement\hspace{0.1em}\textit{adjusting the object distance from the camera} (54\%) and \boxGesture\hspace{0.1em}\textit{pinching} (37\%). Participants often aimed to match the perceived size of the physical object from audiences with the intended scale. For larger objects with broad surfaces (e.g., laptops), pinching gestures were more common, reflecting participants' prior experiences with touchscreen scaling.

\begin{wrapfigure}[2]{l}{0.001\textwidth}
  \begin{center}
    \vspace{-15pt}
    \includegraphics[width=0.035\textwidth]{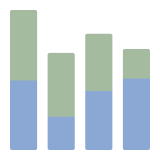}
  \end{center}
\end{wrapfigure}
\noindent\textbf{Composose/Decompose:} 
Most participants mapped this command to \textbf{Arrangement-based Interactions}, particularly \boxCombination\hspace{0.1em}\textit{adjusting the distance between objects} (89\%). The proposed gestures were consistent across all physical objects, intuitively associating the close proximity of two physical objects with the composition and the distant proximity with the decomposition.

\begin{wrapfigure}[2]{l}{0.001\textwidth}
  \begin{center}
    \vspace{-15pt}
    \includegraphics[width=0.035\textwidth]{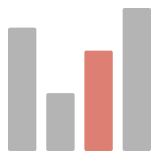}
  \end{center}
\end{wrapfigure}
\noindent\textbf{Select/Deselect Data Points:} Participants favored \textbf{Visualization-based Interactions}, especially \boxDirect\hspace{0.1em}\textit{pointing gestures} (57\%). 
As W3 noted, \textit{``When the visualization is directly in front of me, it feels inefficient to involve a physical object for selecting a single data point.''} Others (e.g., W1) proposed \boxCombination\hspace{0.1em}\textit{isolation} (27\%) for small visual targets, such as points in a scatterplot, noting that selecting tiny elements with fingers can be difficult.


\begin{wrapfigure}[2]{l}{0.001\textwidth}
  \begin{center}
    \vspace{-15pt}
    \includegraphics[width=0.035\textwidth]{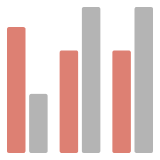}
  \end{center}
\end{wrapfigure}
\noindent\textbf{Select/Deselect Data Series:} 
\textbf{Arrangement-based Interactions}, specifically \boxCombination\hspace{0.1em}\textit{isolation of the target object} (40\%), was the most favored approach.  Isolation involves separating one object spatially from others to select its linked data series. As W1 noted, \textit{``When an object stands apart visually, it makes sense that the audience would focus on it---and the system could then highlight the associated data.''}

\begin{wrapfigure}[2]{l}{0.001\textwidth}
  \begin{center}
    \vspace{-15pt}
    \includegraphics[width=0.035\textwidth]{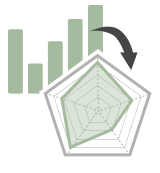}
  \end{center}
\end{wrapfigure}
\noindent\textbf{Change Chart Types:}
No dominant strategy emerged from the participants' responses. Participants proposed \boxVisibility\hspace{0.1em}\textit{changing the object's appearance} (26\%), such as peeling a banana or swapping it with a different-colored object, to signal a change in chart type visually. Others suggested \boxMovement\hspace{0.1em}\textit{rotating the object} (21\%) to switch between visualizations, mapping the physical object's orientation to different chart types.

\begin{wrapfigure}[2]{l}{0.001\textwidth}
  \begin{center}
    \vspace{-15pt}
    \includegraphics[width=0.035\textwidth]{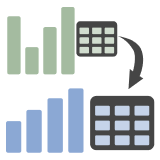}
  \end{center}
\end{wrapfigure}
\noindent\textbf{Change Data Source:} 
Most participants associated this with \textbf{Appearance-based Interactions}, particularly \boxVisibility\hspace{0.1em}\textit{modifying the object's appearance} (61\%). Similar to changing visualization type, these manipulations symbolically conveyed a shift in the underlying data source, such as replacing or revealing new aspects of the object.

\begin{wrapfigure}[2]{l}{0.001\textwidth}
  \begin{center}
    \vspace{-15pt}
    \includegraphics[width=0.035\textwidth]{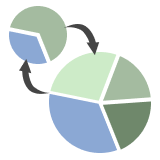}
  \end{center}
\end{wrapfigure}
\noindent\textbf{Hierarchical Navigation:} 
No single strategy dominated. Participants frequently proposed \boxMovement\hspace{0.1em}\textit{adjusting the object's distance from the camera} (36\%), \boxVisibility\hspace{0.1em}\textit{controlling access} (21\%), and \boxGesture\hspace{0.1em}\textit{pinching} (20\%).
While conceptually similar to visual scaling, \textit{controlling access} was often preferred for its stronger semantic alignment with the idea of revealing more detailed content.
As W8 described, \textit{``Opening a backpack to show what's inside just feels like drilling down into the data.''}
\begin{figure*}[ht]
    \centering
    \includegraphics[width=1\linewidth]{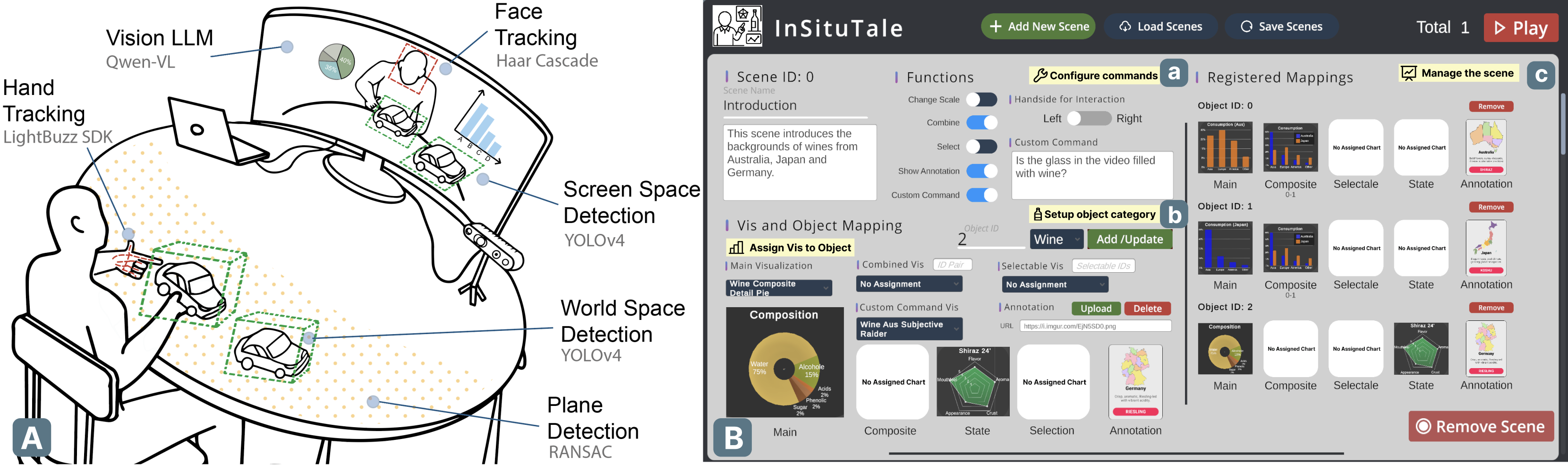}
    \caption{Presentation and Authoring Modes: \textit{InSituTale} consists of presentation and authoring mode.
(A) In presentation mode, the presenter interacts with visualizations in real time using physical objects and hand gestures, tracked by a depth camera placed in front of them. A live video stream is displayed locally for monitoring.
(B) In authoring mode, the presenter configures physical objects, commands, and visualizations to be used during the presentation.
}
    \label{fig:system}
    \Description{Figure 4.1: Two images illustrating the components of the \textit{InSituTale} system. (A) A user is seated in front of a desk, interacting with a camera-mounted tabletop setup. The system analyzes the user’s physical interactions in real time using various sensing and recognition technologies. (B) The authoring interface allows users to add new scenes using buttons at the top of the screen. In the main panel, users can toggle interaction types, select physical objects, and assign corresponding visualizations.}
\end{figure*}

\section{Design Process}
We developed \textit{InSituTale} iteratively, drawing our workshop's results on the natural mappings between physical manipulations and visualization commands from our workshop. Throughout this iterative design process, we identified five critical design considerations for  augmented physical data storytelling:

\noindent\circled{D1}\textbf{ Physical Space Interaction Detection:}
Tracking physical objects and presenter gestures in 3D physical space to ensure reliable interaction detection.

\noindent\circled{D2}\textbf{ Dynamic Visualization Placement:}
Automatically positioning visualizations to prevent them from occluding a presenter and physical objects.

\noindent\circled{D3}\textbf{ Minimization of Interaction Ambiguity:}
Providing robust mappings and a minimal set of commands tailored to each storytelling scenario to prevent unintended interactions.

\noindent\circled{D4}\textbf{ Smooth Storytelling Flow:}
Balancing manual and automated visualization controls to support improvisational presentation while minimizing repetitive command execution that may disrupt storytelling flow.

\noindent\circled{D5}\textbf{ Context-Aware Presenter Guidance:}
Providing real-time, scene-specific cues to help presenters remember available interactions and maintain narrative coherence.

\subsection{Iterative Design}
We iteratively refined \textit{InSituTale}'s interaction mechanisms to enhance the overall presentation experience. The prototypes developed during the design process were tested internally by the authors and informally evaluated by external participants.

\subsubsection*{\textbf{First Iteration: Supporting 3D Spatial-Aware Object Tracking.}}
Initially, we used ArUco markers and a webcam to detect manipulations, such as moving a physical object closer to the camera or another object. 
Early testing showed that due to reliance on 2D screen-space tracking, these manipulations involving depth were frequently misinterpreted as horizontal or vertical screen movements, leading to inaccuracies \circled{D1}. Additionally, the markers were often unintentionally obscured by users, causing interaction failures
These limitations motivated our shift to a markerless tracking solution with a depth camera.

\subsubsection*{\textbf{Second Iteration: Improving Interaction Robustness and Flow.}}
We integrated a depth camera with an object detection model, improving tracking accuracy and enabling reliable detection of interactions in physical space \circled{D1}. 
Guided by workshop findings, we implemented common mappings between physical manipulations and visualization commands while ensuring that each manipulation remained distinct. 
Early user tests revealed that requiring presenters to manually execute frequently used commands, such as showing or hiding visualizations, disrupted the storytelling flow due to repetition. 
To mitigate this, we introduced a scene-based presentation structure, allowing presenters to assign visualizations to individual scenes and automatically control their visibility through scene transitions inspired by prior work~\cite{chironomia} \circled{D4}. 
Presenters can navigate scenes both forward and backward as needed.
\ac{While this structure imposes some constraints on the overall sequence of scenes, it still offers flexibility within each scene. Presenters can vary the order of interactions and repeat actions as needed.}
Further testing also showed frequent misclassification of gestures, particularly between pointing and lifting. 
Since presenters often use their index fingers while grasping objects, these interactions were occasionally misrecognized as pointing. 
To mitigate this, we allowed presenters to specify a preferred pointing hand (left or right), ensuring that gestures made with the non-dominant hand would not be interpreted as pointing \circled{D3}.
Additionally, we enabled presenters to selectively activate or deactivate visualization commands on a per-scene basis to reduce the likelihood of unintended triggers \circled{D3}. 
This approach reflects a balance between reducing false detections and supporting improvisation.

\subsubsection*{\textbf{Third Iteration: Supporting Presenters with Contextual Awareness.}
}
We introduced support for multiple scenes, each with individually configurable commands and visualizations. Internal tests revealed that presenters frequently forgot which visualizations and interactions were associated with each scene, creating uncertainty during presentations. To overcome this issue, we added an on-screen guidance panel---visible only to presenters---to display configured interactions and associated visualizations. This panel substantially reduced cognitive load by providing immediate and contextual assistance \circled{D5}.

\begin{figure*}[ht]
    \centering
    \includegraphics[width=1\linewidth]{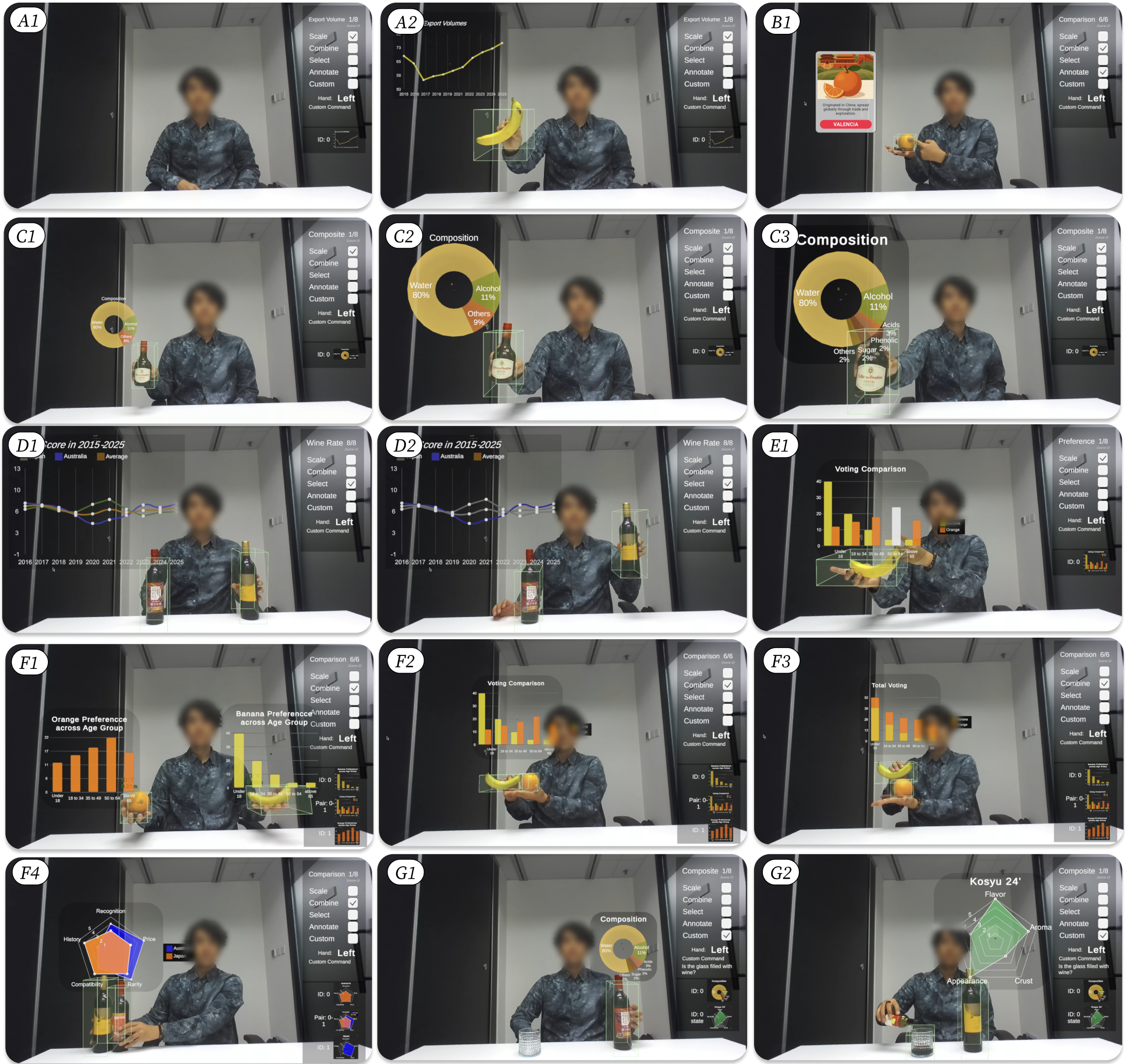}
    \caption{
Supported Interactions: (A) Place an object to show a chart.
(B) Point at an object using the index finger to show an annotation.
(C) Move an object closer to the camera to scale the visualization and reveal details.
(D) Lift an object to highlight the associated data series.
(E) Point to a data point using the index finger.
(F) Bring objects closer together to generate a composite chart.
(G) Trigger a visualization change via a user-defined condition. Bounding boxes and the presentation panel are not shown to the audience.}
    \label{fig:InSituTaleSystemCapture}
    \Description{Figure 5.1
A multi-part image showing supported interactions in \textit{InSituTale}. (A1–A2) A chart is displayed when a physical object is placed in view of the camera. (B1) Pointing at an object with a registered index finger reveals an annotation. (C1–C2) Lifting the object highlights a corresponding data series. (D1) Pointing at a specific data point highlights that point. (E1–E3) Moving the object closer to the camera enlarges the visualization, revealing more detail. (F1–F4) Bringing multiple objects together creates a composite visualization; the layout adapts to the alignment, forming either a stacked or clustered bar chart. (G1–G2) Performing a user-defined condition, such as holding a wine-filled glass, triggers a transition to an alternative visualization.
}
\end{figure*}

\section{InSituTale}
This section presents the final design of \textit{InSituTale}.
\textit{InSituTale} comprises two modes: Presentation Mode (\cref{fig:system}-A), which enables real-time storytelling through physical interactions, and Authoring Mode (\cref{fig:system}-B), allowing presenters to configure scenes, visualizations, and interactions beforehand.

\subsection{Presentation Mode}
\textit{InSituTale} supports real-time data storytelling through interaction with physical objects on a table. A depth camera captures the presenter's physical space.
The system generates a live video stream that combines physical footage (i.e., the presenter and objects) with augmented visualizations, viewable by both presenters and remote audiences.

\subsubsection{Visualization Visibility and Scaling}
Presenters can dynamically control the visibility and scale of visualizations. Placing a physical object within the camera view reveals the corresponding visualization\ac{, while hiding it can be achieved either by moving the object out of the camera’s view or by shifting it beyond a predefined distance threshold, as detected by a depth camera (\cref{fig:InSituTaleSystemCapture}-A1,A2).}
Presenters can display annotations (images or text) by pointing at a physical object with their index finger for a preset duration (\cref{fig:InSituTaleSystemCapture}-B1). Moving an object closer to or farther from the camera scales the visualization accordingly.
Additionally, placing an object closer to the camera than a distance threshold triggers a transition to a more detailed view (\cref{fig:InSituTaleSystemCapture}-C1--C3).

\subsubsection{Data Selection}
Presenters can highlight specific data series linked to physical objects. A data series is selected by lifting an object from the table while keeping others stationary (\cref{fig:InSituTaleSystemCapture}-D1, D2).
Individual data points can also be highlighted by pointing directly at the visualization with an index finger (\cref{fig:InSituTaleSystemCapture}-E1).

\subsubsection{Visualization Composition and Transformation}
Presenters can dynamically create composite visualizations or transform existing ones into different types. Composite visualizations are generated when multiple related objects are brought close together horizontally or vertically. 
For bar charts, horizontal alignment results in a clustered bar chart, while vertical alignment produces a stacked bar chart (\cref{fig:InSituTaleSystemCapture}-F1--F3). 
Other chart types---including pie, donut, radar, and line charts---are composited through visual overlay, regardless of their spatial orientation (\cref{fig:InSituTaleSystemCapture}-F4). 
The system can also trigger visualization transformations based on detected real-world events. 
For instance, a presenter may define a condition such as \textit{``Is the glass filled with red wine?''}. 
When the condition is met, the system automatically replaces the associated visualizations with pre-registered ones (\cref{fig:InSituTaleSystemCapture}-G1,G2).

\subsubsection{Scene Control}
Presenters navigate through predefined scenes, each containing unique visualizations and enabled interaction commands. They can move forward or backward between scenes as needed.
Scene transitions are controlled via a mouse right-click, assuming presenters will use a commonly adopted presentation tool, such as a ring mouse or clicker.
This design choice is informed by insights from existing synchronous data storytelling systems~\cite{chironomia}, ensuring that scene transitions remain distinct from performative storytelling gestures while maintaining a smooth and controlled presentation flow. 

\subsubsection{Dynamic Visual Arrangement}
\textit{InSituTale} dynamically positions visualizations near their corresponding physical objects, ensuring that data remains contextually linked to the objects being referenced.
When presenters move objects, visualizations update their positions in real time. The system employs an adaptive layout algorithm to prevent overlaps between visualizations, physical objects, and the presenter's face. Composite visualizations are centered above the contributing physical objects, and visualizations are constrained within a display boundary to prevent unintended cropping.1

\subsubsection{Presenter Support}
To reduce cognitive load and potential misuse of commands, \textit{InSituTale} provides presenters with a private interface that displays key contextual information for each scene (see right panels in \cref{fig:InSituTaleSystemCapture}). 
This interface presents mappings between physical objects and their associated visualizations, a list of active visualization commands, and visualizations registered for transformation and composition.

\subsection{Authoring Mode}
\textit{InSituTale} allows presenters to define scenes, specify target physical objects, assign associated visualizations, and select interaction commands for their presentations. \ac{To support iterative testing during authoring and ensure reproducibility, \textit{InSituTale} also allows users to import templates and edit previously saved settings. }
The authoring interface is organized into three panels: the Interaction Setup panel (\cref{fig:system}\rc{-B-a}), the Visualization Setup panel (\cref{fig:system}\rc{-B-b}), and the Object--Visualization Mapping panel (\cref{fig:system}\rc{-B-c}).
The authoring process begins with structuring the presentation into a sequence of scenes. 
Each scene includes a set of associated visualizations, physical objects, and visualization commands. Presenters navigate through these scenes during the live presentation, with the flexibility to move forward or backward as needed. 
In the Interaction Setup panel (\cref{fig:system}\rc{-B-a}), presenters enable specific visualization commands and select the preferred hand (left or right) for pointing gestures. They can also define text-based conditions to trigger visualization transformations (e.g., \textit{``Is the glass filled with red wine?''}). In the Visualization Setup panel (\cref{fig:system}\rc{-B-b}), presenters select physical objects from a predefined class set and assign each object a unique identifier for runtime tracking. While \textit{InSituTale} does not support custom visualization authoring, presenters can link physical objects to preconfigured Unity visualization prefabs (e.g., pie, bar, line, and radar charts). Annotation images and texts can also be uploaded via a web interface. All mappings and configurations are summarized in the Object–Visualization Mapping panel (\cref{fig:system}\rc{-B-c}) for easy review and management.

\begin{figure}[t]
    \centering
    \includegraphics[width=0.95\linewidth]{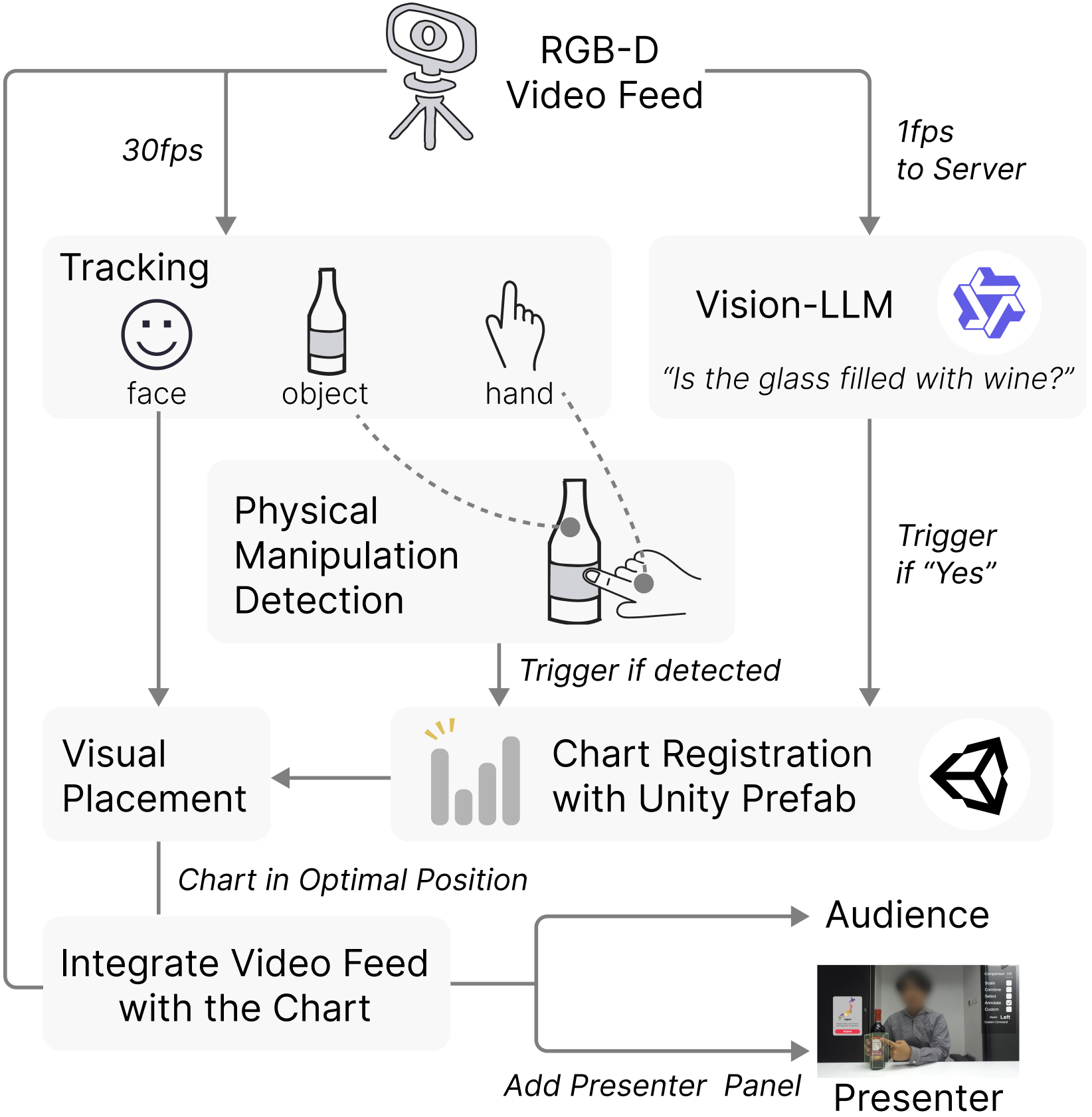}
     \caption{System Flow: \textit{InSituTale} consists of object, hand, and face tracking, Vision-LLM, and visualization placement optimization.}
    \label{fig:systemflow}
    \Description{.}
\end{figure}

\begin{figure*}[ht]
    \centering
    \includegraphics[width=1\linewidth]{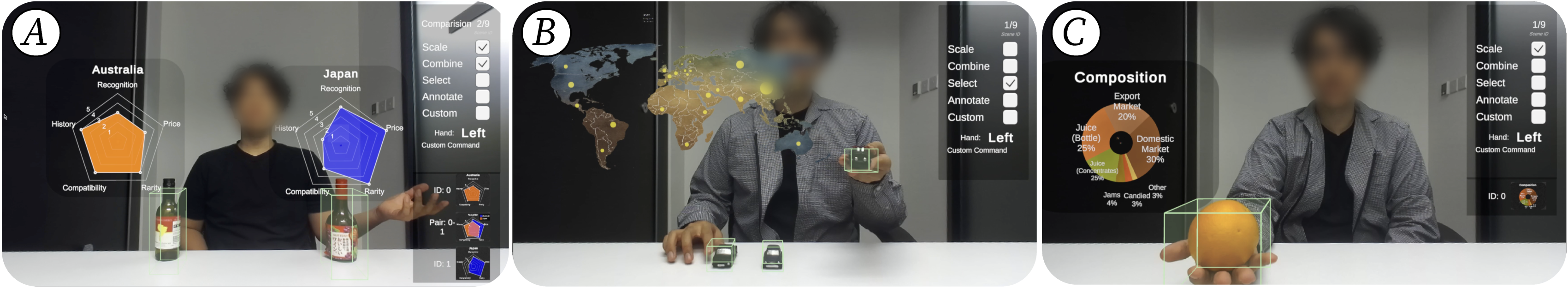}
    \caption{Use Cases: (A) Comparing the characteristics of Japanese and Australian wines. (B) Highlighting the global market share of EV models. (C) Explaining the detailed consumption breakdown of oranges.}
    \label{fig:usecase}
    \Description{
Figure 6.1
A composite image illustrating three use cases of the InSituTale system. In (A), a presenter brings two wine bottles closer together on a tabletop, triggering the display of a clustered bar chart that enables easier visual comparison between the two items. In (B), the presenter lifts a physical model of an electric vehicle, causing the system to highlight corresponding data points in a chart, drawing attention to the selected item. In (C), an orange is moved closer to the camera, prompting the visualization to switch to a more detailed view of a pie chart related to that object. Each use case demonstrates how physical object manipulation is used to control data visualizations in real time.}
\end{figure*}

\subsection{Implementation}
We implemented \textit{InSituTale} in Unity. The system captures the physical environment using a depth camera, analyzes the video feed in real time, and augments the physical environment with visualizations.  \cref{fig:systemflow} illustrates the overall system flow.

\subsubsection{Object Tracking} 
\label{sec:object-tracking}
\textit{InSituTale} captures RGB-D video streams, enabling simultaneous object recognition and spatial tracking in 3D space. We employ a YOLOv4-based object detection model trained on 80 categories from the COCO dataset~\cite{cocodataset}, allowing class-level detection of physical objects. The detection model can be replaced to accommodate domain-specific object types. Detected objects are classified at the class level and assigned temporary IDs based on the order of detection within each frame. Persistent tracking across frames is achieved via a heuristic matching algorithm that associates new detections with previously tracked instances based on spatial proximity and positional continuity. To detect Compose/Decompose interactions, the system identifies the relative distances between physical objects. The system also employs plane detection to establish a baseline surface, allowing the recognition of object-lifting gestures by monitoring vertical displacement from the baseline surface. Additionally, the visualization scaling is dynamically calculated based on the distance between objects and the camera.

\subsubsection{Query-Based Recognition} To enable the detection of real-world state changes aligned with narrative contexts, we incorporated a large vision-language model for real-time video analysis. We employed Qwen-VL-Chat~\cite{Qwen-VL}, an open-source, instruction-tuned vision-language model. Each second, the system processes one compressed video frame via socket communication to this model. Frames are analyzed alongside tailored textual prompts (e.g., \textit{``Is the glass filled with red wine? Respond 1 if yes, 0 if no.''}) to detect context-specific events defined during authoring.

\subsubsection{Hand, Finger, and Face Tracking} Hand and finger tracking are implemented using the LightBuzz Body Tracking SDK.\footnote{\url{https://handtracking.lightbuzz.com}} The system distinguishes left and right hands and continuously tracks the position of the index finger on the registered side in 2D screen space. By correlating the index finger position with detected objects and visualizations, \textit{InSituTale} recognizes pointing gestures. A face area is also detected using the Haar Cascade method~\cite{haarCascade}, allowing a layout algorithm to prevent visualizations from occluding the presenter's face.

\subsubsection{Visualization Rendering and Layout} \textit{InSituTale} supports both 2D screen space and 3D physical space visualizations, implemented using Graph And Chart Unity Assets\footnote{\url{https://assetstore.unity.com/packages/tools/gui/graph-and-chart-data-visualization-78488}}. 
To ensure visualizations and annotations remain clearly visible without occlusions, we employ a greedy algorithm informed by layout optimization methods~\cite{labelPlacement2003, imageLabelPlacement2012}. 
The algorithm evaluates eight candidate positions around a physical object with a tailored objective function.
This function penalizes overlaps with the presenter's face, other objects, and existing visualizations while rewarding positions aligned with the top side of the object and positions consistent with prior frames to avoid frequent and radical movements of visualizations. 
The candidate position achieving the highest objective score is selected. Position smoothing through linear interpolation further stabilizes visualization movements.

\section{Use Cases}
 \textit{InSituTale} supports a variety of augmented physical data storytelling scenarios (\cref{fig:usecase}). 
 We illustrate three representative examples (see the supplemental video).

\subsection{Promotion Presentation}
In a virtual wine-tasting session, a sommelier uses \textit{InSituTale} to introduce and compare two wines  (\cref{fig:usecase}-A). Placing the first bottle on the table triggers a radar chart that visualizes key characteristics such as brand recognition, food pairing compatibility, and pricing. Pointing at the bottle helps orient the audience's attention and reveals an annotation featuring a photo of the vineyard and a brief historical description. 
The second wine is placed similarly with its own radar chart and annotation. \textit{InSituTale} automatically arranges the visuals to avoid occlusion with the presenter or other elements in the scene. To compare the wines, the sommelier brings the two bottles close together, prompting the system to overlay their radar charts and highlight similarities and differences (\cref{fig:InSituTaleSystemCapture}-F4). In the next scene, advanced via a ring mouse, the system shows a multi-series line chart showing evaluation scores across vintages for both wines alongside an industry benchmark (\cref{fig:InSituTaleSystemCapture}-D1). 
 After explaining overall trends, the sommelier responds to a viewer's request by improvisationally highlighting the line corresponding to the Australian wine (\cref{fig:InSituTaleSystemCapture}-D2), making it easier for the audience to follow. Finally, the sommelier opens one of the bottles and pours wine into a glass. This action, linked to the system prompt \textit{``Is the glass filled with wine?''}, triggers a new radar chart displaying detailed tasting notes, including aroma, flavor, and texture (\cref{fig:InSituTaleSystemCapture}-G2), offering the audience a richer and more immersive tasting experience.

\subsection{Product Comparison Presentation}
In a virtual product showcase, a presenter compares electric vehicles (EVs), hybrid cars, and gasoline-powered vehicles using small-scale car models (\cref{fig:usecase}-B). The presentation begins with a world map where dot sizes represent sales volumes for each vehicle type. 
The presenter first lifts the EV model, directing the audience's attention to that category and simultaneously highlighting the corresponding dots on the map. In the next scene, two bar charts appear side by side, showing regional sales for EVs and hybrids. Moving the vehicle models horizontally causes the charts to merge into a clustered bar chart for direct comparison. Stacking the models vertically transforms the visualization into a stacked bar chart, representing the combined market share and illustrating shifts from gasoline to electric drivetrains. In the next scene, placing each model individually triggers a pie chart that breaks down cost components such as manufacturing, battery, maintenance, and fuel. Bringing the EV model closer to the camera enlarges the chart and reveals finer segments, including details like battery sourcing and warranty coverage, enabling a more nuanced cost analysis.

\subsection{Consumer Behavior Presentation}
In a presentation on global fruit consumption, a speaker uses an orange and a banana to illustrate market trends (\cref{fig:usecase}-C). Placing the orange on the table triggers a pie chart showing its consumption breakdown—fresh use, juice production, and processed products. Similarly, placing the banana reveals its usage distribution, such as in smoothies, baby food, and fresh consumption. 
Pointing at each fruit brings up annotations: for example, the banana's annotation shows an image and caption highlighting its use in school lunch programs and the logistical advantages of its protective peel, which reduces bruising and waste. Later in the presentation, the speaker peels the banana in front of the camera (\cref{fig:teaser}). This action, recognized by the prompt \textit{``Is the banana peeled?''}, updates the annotation with information about ripeness levels and glycemic index, illustrating how sensory and nutritional properties change as the fruit matures.
\section{Evaluation}
We conducted a user study to evaluate how presenters experience delivering data stories using \textit{InSituTale}, focusing on its usability, utility, and learnability in real-time storytelling contexts.

\subsection{Participants}
We recruited 12 participants (P1--P12) with diverse backgrounds, including six males and six females, ranging in age from 19 to 30. Most had experience delivering data-centric presentations, though their familiarity with augmented reality varied from none to extensive. All participants attended the study in person.

\subsection{Apparatus}
The system ran on a laptop equipped with an Intel Core i7 processor (3.6 GHz), 32 GB RAM, and an NVIDIA RTX 3070 GPU. 
A separate RTX 3090 GPU hosted the vision-language model. 
A ZED Mini stereo camera, mounted on a tripod 50 cm in front of the presenter, captured RGB-D video input. Participants sat at a table with physical props placed in front of them. The instructor monitored the system output on a remote display.
\ac{With this setup, the average latency from frame capture to visual change was 0.1 seconds ($SD = 0.02$), and the response time from the LLM server averaged 1.08 seconds ($SD = 0.067$).}

\subsection{Tasks and Procedure}
Participants were asked to use \textit{InSituTale} to deliver a data story in which physical objects served as key narrative props. We prepared two story sets. One set focused on \textbf{Japanese and Australian wine} and included two wine bottles and a wine glass as props. The dataset contained information such as the typical composition of wines from each region (pie chart), consumption statistics across global regions (bar chart), regional market characteristics (radar chart), tasting notes (radar chart), average ratings over the past decade (line chart), and annotated images illustrating the distinctive character of each country's wines. The second story featured a comparison between \textbf{bananas and oranges}, supported by two of each fruit as physical props. This content included market share breakdowns (donut chart), preference survey results by age group (bar chart), attribute comparisons (radar chart), export volumes (line chart), retail distribution channels (bar chart), and annotated images of the fruits' countries of origin. Participants were free to determine the structure and focus of their presentations; they were not required to use all the provided materials or follow a specific order. Instead, they were encouraged to construct a coherent narrative using a subset of the content based on their preferences.

The study was comprised of three phases: training, authoring, and presentation, lasting approximately 50--60 minutes in total. In the training phase (approximately 10 minutes), participants completed a consent form and demographic questionnaire and received a guided demonstration of \textit{InSituTale}'s core features. During the authoring phase (15--25 minutes), each participant was assigned one of the story sets. They received the relevant physical objects and a printed handout summarizing the dataset.  
Using the authoring interface, participants configured scenes by assigning visualizations to physical objects and enabling interaction commands. They were allowed to ask for support from the instructor as needed, including helping them configure additional visualizations. Participants also searched for annotation images online and uploaded them through the system interface if desired.
In the presentation phase (10--15 minutes), participants delivered their authored presentations using \textit{InSituTale}. 
A ring mouse was used to trigger scene transitions. 
The instructor remained available but answered questions only upon request. After completing their presentations, participants filled out a usability questionnaire. They took part in a semi-structured interview to share qualitative feedback on their experience, the system's usability, and any perceived limitations.

\subsection{Results}
We analyzed the responses from the usability questionnaires and semi-structured interviews. A summary of the usability ratings is shown in \cref{fig:resultPresentation}.

\subsubsection*{\textbf{Perceived Value and Benefits}}
\begin{figure}[t]
    \centering
    \includegraphics[width=1\linewidth]{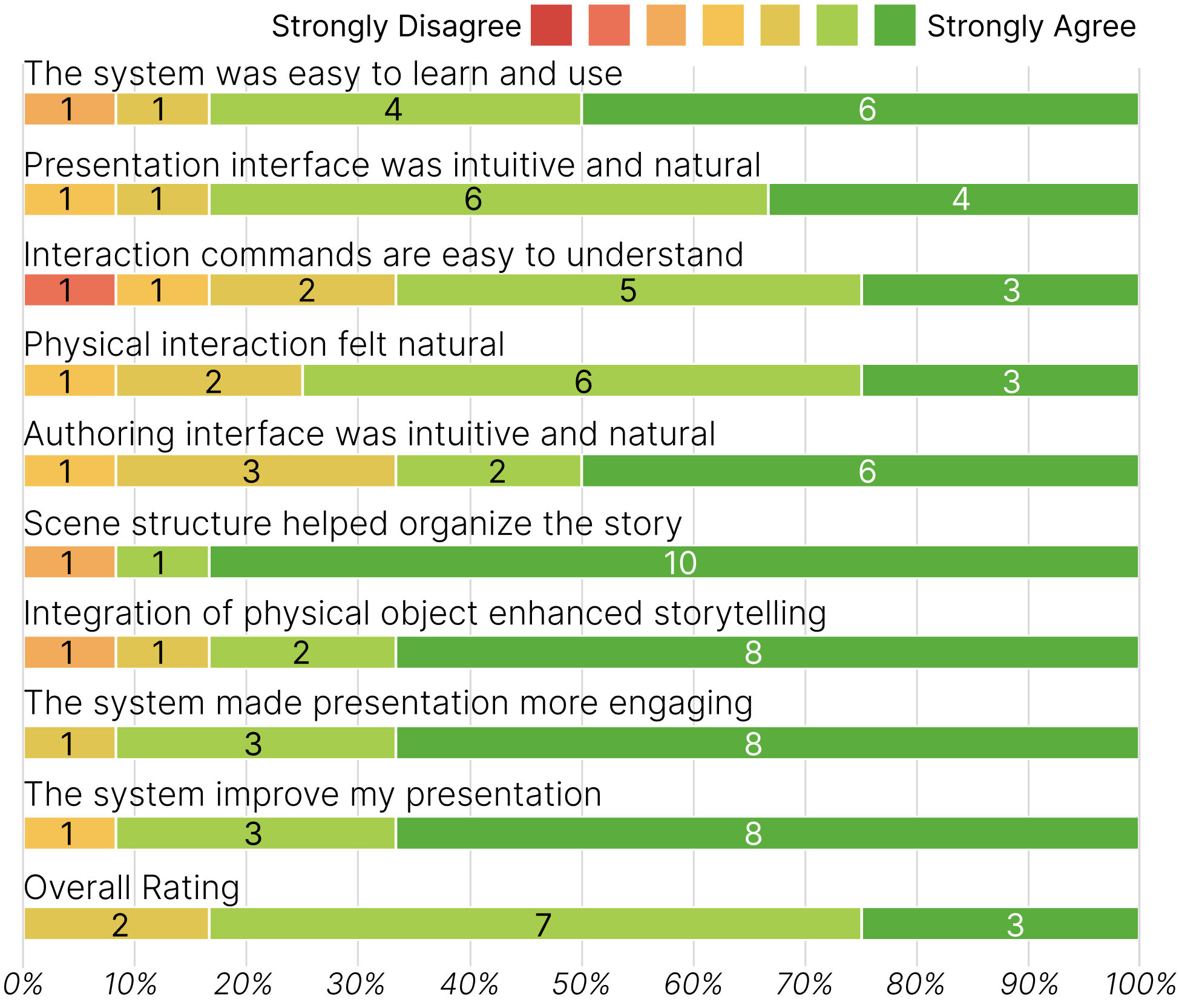}
    \caption{Usability Ratings: Participants rated the usability of \textit{InSituTale} on a 7-point Likert scale.}
    \label{fig:resultPresentation}
    \Description{Figure 6.1
This figure presents a horizontal stacked bar chart that visualizes participant responses to eleven evaluation statements regarding the InSituTale system. Each response was rated on a 7-point Likert scale ranging from ``Strongly Disagree'' (1, shown in dark red) to ``Strongly Agree'' (7, shown in dark green), with intermediate levels represented by gradually shifting color tones. The chart shows that most participants agreed strongly that the system was easy to learn and use, with over half of the responses at levels 6 and 7. Similarly, the presentation interface was rated as intuitive and natural, with nearly all responses clustered in the highest agreement levels. The interaction commands were generally considered easy to understand, though there was a small portion of responses at lower levels, indicating that one or two participants found them less straightforward. Participants also responded positively to the naturalness of physical interaction, and the majority indicated strong agreement that the integration of physical objects enhanced storytelling. The authoring interface and the scene structure were also well-received, although responses here showed slightly more variation across mid-to-high agreement levels. In terms of engagement, most users felt that the system made their presentations more engaging. While the perception of whether it improved their presentation varied slightly, the overall trend was still positive. Finally, the overall rating for the system showed strong support, with most responses again in the highest range of agreement. The x-axis represents percentage from 0\% to 100\%, allowing easy comparison of agreement levels across the different statements.}
\end{figure}

Participants recognized the clear value in using \textit{InSituTale} to enhance their current presentation practices. A majority (11 out of 12) agreed that the system improved their overall storytelling experience (\cref{fig:resultPresentation}).
P1, reflecting on their experience compared to traditional slideshow-based presentations, noted: \textit{``I liked how I could improvise with visualizations during the presentation. I ended up highlighting data points and using composite charts that I hadn't originally planned. This isn't possible in regular slides.''} The ability to modify visualizations in real time, along with physical object manipulations, offered presenters a more dynamic and personalized storytelling experience. Participants also appreciated the integrated view of physical objects and visualizations. P1 shared: \textit{``When explaining granite or basalt, professors show slides alongside physical samples. With InSituTale, they could directly link information to the objects, helping students focus on the materials.''} Such integrated views can effectively support storytelling in scenarios where physical objects function as essential narrative props. 
Moreover, participants highlighted how the system handled visual placement. Unlike conventional video conferencing platforms like Zoom, which divide the screen between the speaker and visual aids, \textit{InSituTale} allows visualizations, physical props, and the presenter to coexist in a single unified frame. P4 commented: \textit{``I liked how the visualizations automatically moved to avoid covering my face or other objects. It let me stay focused on presenting rather than worrying about where visuals would appear.''}

On the other hand, some participants expressed a desire for greater expressiveness. P2, who frequently gives data-driven presentations, noted: \textit{``I expected to zoom into specific sections of bar charts, but that wasn't supported,''} highlighting the need for more fine-grained control over visualizations. P3 suggested incorporating 3D models for comparison, reflecting interest in a broader range of visual elements. P7 and P8 proposed dynamic annotations (similar to pen tools in PowerPoint) that could persistently highlight areas of interest. 
P2 also envisioned a higher degree of physical-digital coordination:
\textit{``It would be interesting to use a robot to move physical objects in sync with the visualizations.''} 
Similarly, P10 imagined visual properties of physical objects adapting in real time:\textit{``I'd like to change the appearance of a physical object, like altering a label’s color, to match the visualization.''}

\subsubsection*{\textbf{Engagement and Interaction Experience}}
All participants agreed that \textit{InSituTale} provides an engaging presentation experience (\cref{fig:resultPresentation}). As P7 remarked, \textit{``I enjoyed the physical interactions---it felt more like a performance than just clicking through slides.''}
Many participants also appreciated the customizable object detection and actively experimented with their own prompts (e.g., \textit{``Is the glass filled with wine?''}, \textit{``Is the presenter dancing?''}, \textit{``Is the presenter eating an orange?''}). P1 noted, \textit{``I really liked being able to define my own triggers. It was fun when a wine-drinking gesture activated a chart. I want to explore it more.''}
Participants generally found the interaction techniques easy to learn. Most picked them up after a single demonstration. P8 explained, \textit{``During my wine presentation, the system's responses to my gestures with wine bottles felt intuitive. I could remember the mappings easily because they made sense.''}
Furthermore, having visualizations automatically respond to physical object interactions helped maintain a smooth presentation flow. As P5 noted, \textit{``Picking up, moving closer, or pointing at objects---things I'd do anyway while explaining---automatically triggered the visuals.''} Several participants also reported less physical fatigue compared to gesture-based systems. Drawing on prior VR experience, P7 remarked, \textit{``Using physical objects feels much less tiring than waving your hands in the air.''}

However, a few participants noted unintended interactions. For example, P4 accidentally triggered an annotation while lifting an object due to a misinterpreted pointing gesture: \textit{``It's hard to keep track of which hand is being used for pointing while you're presenting.''} Some objects were also harder to detect depending on how they were held. P6 noted, \textit{``The orange worked fine, but holding the banana the same way as the orange blocked detecting the banana, so I had to change my grip.''} 
Additionally, several participants (e.g., P11) expressed interest in integrating voice commands to further extend the interaction mechanism.

\subsubsection*{\textbf{Scene Management and System Behavior}}
Participants generally responded positively to \textit{InSituTale}’s scene-based structure. Most participants created 4--6 scenes, typically aligning one scene per data point or topic. As P3 noted, \textit{``It felt similar to making slides---you assign each thing you want to say to its own scene.''} The scene-specific presentation panel was also well-received, with participants agreeing that it supported smoother delivery. P7 shared, \textit{``Having the panel was reassuring---I didn’t need to remember which interaction triggered which visualization.''}

However, some participants noted limitations in how the system handled object-to-visualization mappings---particularly when multiple objects from the same class were used. The system assigns visualizations based on the detection order, which can be unintuitive during improvisational use. P4 remarked, \textit{``The panel shows which visualization comes first, but being restricted by this sequence reduces the improvisational strength.''}
This limitation was particularly noticeable when object tracking failed mid-presentation, causing visualizations to mismatch. 
P5 and P6 described needing to remove all objects from the scene and reintroduce them in the intended order. While P5 noted, \textit{``Fixing the assignments was straightforward,''} the issue nevertheless highlighted a key challenge for real-time, flexible storytelling.

\subsubsection*{\textbf{Authoring Experience}}
Participant reactions to the authoring interface varied. P3 found the scene-based structure intuitive, noting, \textit{``Creating scenes and assigning visualizations felt similar to building slides in PowerPoint.''} 
In contrast, P6 commented, \textit{``Configuring each scene was a bit overwhelming, especially when assigning visualizations triggered by commands (e.g., changing chart types).''} 
Additionally, several participants framed vision-language prompts from a first-person perspective (e.g., \textit{``Am I peeling the banana?''}), which reduced detection accuracy. The system performed more reliably with third-person phrasings, such as \textit{``Is the person (in the scene) peeling the banana?''}, highlighting the need for clearer authoring guidance for creating prompts. 

\subsubsection*{\textbf{Potential Usage Scenarios}}
Participants proposed diverse use cases based on their backgrounds. P1 stated, \textit{``In chemistry demonstrations, it could effectively show how combining materials alters their properties.''} P7 highlighted its suitability for remote sales presentations, noting, \textit{``It focuses customers' attention on the actual products, unlike traditional slides. Sales conversations often require adapting the presentation spontaneously.''} Conversely, some participants noted limitations in co-located scenarios where the presenter and audience share the same space. P5 emphasized, \textit{``Ultimately, product promotion requires letting customers physically handle the products. Therefore, visualization overlays should support co-located contexts, not just remote presentations.''} 
\section{Discussion}
\label{sec:discussion}
We summarize design implications and potential improvements from our design process and user study.

\subsection{Design Implication}
\subsubsection*{\textbf{Semantic Coherence Between Objects and Visuals}}
Aligning the semantic relationships between physical objects and their associated visualizations offers multiple benefits. Beyond enhancing visual coherence for the audience, these correspondences make interactions more intuitive and predictable---presenters can anticipate the effects of physical manipulations based on each object's narrative role. 
Physical objects also serve as expressive intermediaries, affording a wider range of interactions than hand gestures alone, especially for complex or fine-grained commands. Future systems could further strengthen this coherence through bidirectional coupling between the physical and digital. For example, changes in the visualization could trigger updates in the physical environment (e.g., repositioning a physical object when a data point is selected), fostering a more fluid, engaging, and semantically rich storytelling experience.

\subsubsection*{\textbf{Object-Specific Interactions}}
Our workshop revealed a wide variety of proposed physical manipulations, many of which were strongly tied to the affordances of specific physical objects. 
Supporting this diversity requires systems to account for both object-specific interpretations and more general physical manipulations that apply across a range of objects---ensuring flexibility without constraining the system to particular object types. 
To address this, we implemented mappings based on commonly proposed physical manipulations and integrated a vision-language model that allows presenters to define custom queries reflecting physical state changes. This feature was positively received, with participants appreciating the ability to create interactions beyond predefined gesture sets. 
However, we observed limitations in the system’s sensitivity to ambiguous visual changes, as well as challenges for users in formulating effective textual queries during authoring. Future systems could improve support by providing intelligent interfaces that help presenters express intended real-world conditions in ways that are reliably detectable.

\subsubsection*{\textbf{Dynamic Visualization Placement for Enhanced Visibility}}
Effective visualization placement is essential in dynamic presentation settings where physical elements---including the size, number, and positions of objects and presenters---continuously change. Throughout prototyping and evaluation, we observed that poor alignment often led to the occlusion of key content or disrupted audience comprehension. 
Our dynamic layout algorithm mitigated some of these issues by minimizing overlaps and adjusting positions in real-time. 
However, visual clarity can be further improved through adaptive rendering strategies---such as changing chart colors, labels, or transparency---based on object characteristics, layout constraints, or surrounding content~\cite{dataTour, labelPlacement}. As augmented physical data storytelling moves into diverse environments (e.g., classrooms and public exhibitions), ensuring persistent visibility and coherence will be increasingly essential.

\subsubsection*{\textbf{Robustness and Flexibility in Object Tracking}}
Our vision-based object detection pipeline provided stable tracking without the need for external electronics. However, its reliability occasionally suffered due to user handling variations (e.g., rapid or partial object movements). A key limitation arose when multiple objects of the same class were used: the heuristic ID assignment based on detection order constrained improvisational flexibility. To address this, future systems could explore hybrid tracking strategies---such as combining markerless methods with lightweight markers---to achieve more robust object differentiation while preserving usability and visual simplicity.

\subsection{Limitation and Future Work}
\subsubsection*{\textbf{\ac{Extending Interaction Capabilities}}}
\textit{InSituTale} \rc{currently supports physical interactions, but future versions could benefit from enhanced multimodal capabilities to improve expressiveness and adaptability. 
Prior work has shown that voice and gesture-based inputs can enrich real-time presentations~\cite{chironomia, realityTalk}. 
Both workshop participants and study users suggested combining modalities, such as lifting an object while speaking, to enable more natural and flexible interaction patterns.
Incorporating multimodal input would allow presenters to coordinate physical, verbal, and gestural cues in more expressive and intuitive ways.}

\ac{In addition to expanding presenter input, integrating lightweight audience feedback mechanisms could further enhance communication. Monitoring real-time audience reactions remains a challenge, often limiting presenters' ability to respond effectively~\cite{AffectiveSpotlight}. Adding simple feedback channels, such as reaction icons, could help presenters better assess engagement and adjust their storytelling in the moment.}

\subsubsection*{\textbf{\ac{Audience-Oriented Evaluation}}}
\ac{While \textit{InSituTale} targets real-time data presentations, our evaluation focused mainly on presenter experience and system functionality. We did not formally assess how audiences perceive or benefit from InSituTale compared to conventional slide-based presentations, which we see as the most relevant baseline given its focus on real-time delivery. Although some presenters noted improved clarity and engagement, systematic audience-side evaluation, such as measuring comprehension or attention across different contexts, remains important future work.}

\subsubsection*{\textbf{Limited Object Diversity}} Our interaction design was grounded in six representative physical objects with varied affordances. While informative, this range is limited, and the identified interactions may not generalize to objects with different forms or materials.
\ac{Moreover, the semantic mappings are still influenced by the chosen set of objects.}
A broader exploration of physical objects could uncover a wider range of interaction possibilities and improve adaptability across diverse storytelling contexts.

\subsubsection*{\textbf{Focus on Remote Presentations}}
\textit{InSituTale} was primarily developed for remote settings.
However, several participants noted its potential in co-located environments. Future work could explore adapting the system for in-person use with projection-based AR or HMDs, similar to existing study approaches~\cite{TangibleNet, Tangible3DTableTop}. This enables audiences to experience and interact with visualizations and physical objects directly.

\subsubsection*{\textbf{\ac{Authoring Support}}}
\ac{The current authoring supports of \textit{InSituTale} are limited to using templates and previously saved settings.
Participants also noted challenges in associating physical objects with appropriate visuals for each scene and narrative prompt.
To enhance usability, future systems could offer intelligent recommendations and allow users to define outputs through direct manipulation of physical objects~\cite{TeachableReality, AniCraft}, supporting a more intuitive and flexible authoring experience.}

\section{Conclusion}
\label{sec:conclusion}
This study explored augmented physical data storytelling, an approach that enables presenters to control visualizations through intuitive manipulations of physical objects, seamlessly blending the physical and digital.
We conducted workshops with nine VIS/HCI researchers to investigate how different types of physical actions can be mapped to visualization behaviors in narrative contexts. These insights informed the design of \textit{InSituTale}, a prototype system that integrates physical-space object tracking via a depth camera, a vision-language model for customized detection, and dynamic visualization placement.
Our evaluation with 12 participants showed that \textit{InSituTale}  enables intuitive, engaging, and expressive data storytelling through cohesive integration of physical and digital elements. We hope this work encourages further exploration of physical-object-based interaction in data-driven presentation systems.

\begin{acks}
We thank the reviewers for their valuable feedback. We are also especially grateful to Leni Yang at INRIA Bordeaux for the insightful discussions. This work was supported by the Hong Kong Research Grants Council (RGC) General Research Fund (GRF) grant 16214623, the Knut and Alice Wallenberg Foundation under Grant KAW 2019.0024, and JST PRESTO Grant Number JPMJPR23I5. Takanori Fujiwara completed this work while affiliated with Link\"{o}ping University.
\end{acks}

\bibliographystyle{ACM-Reference-Format}
\bibliography{reference}

\end{document}